\documentclass[usenatbib,usegraphicx,usedcolumn]{mn2e}
\usepackage{bm}
\usepackage{amssymb,amsmath}
\usepackage{graphicx}
\usepackage{natbib}
\usepackage{program}
\usepackage{color}
\definecolor{darkgreen}{rgb}{0.15,0.5,0.15}
\definecolor{darkblue}{rgb}{0.15,0.15,0.5}

\newcommand{\kpc}{{\rm kpc}}

\usepackage[backref,breaklinks,colorlinks,citecolor=blue]{hyperref}

\title[SFDM: helping or hurting small-scale problems?]{Scalar Field Dark Matter: Helping or Hurting Small-Scale Problems in Cosmology?}

\author[V.~H.~Robles, J.~S.~Bullock and M.~Boylan-Kolchin]{Victor H. Robles\thanks{E-mail: roblessv@uci.edu}$^{1}$, James S. Bullock$^{1}$, Michael Boylan-Kolchin$^2$ \\
$1$ Department of Physics and Astronomy, University of
       California, Irvine, 4129 Frederick Reines Hall, Irvine, CA 92697,
       USA\\
$2$ Department of Astronomy, The University of Texas at Austin, 2515 Speedway, Stop C1400, Austin, TX 78712-1205, USA       }

\begin{document}
\date{MNRAS, xxx 2018}

\pagerange{\pageref{firstpage}--\pageref{lastpage}} \pubyear{2018}

\maketitle

\label{firstpage}

\begin{abstract}
Building upon results of cosmological simulations of ultra-light scalar field dark matter (SFDM), we present a comprehensive model for the density profiles of SFDM haloes as a function of halo virial mass $M_{\rm h}$ and scalar field mass $m$.   The central regions of SFDM haloes are dominated by solitons with characteristic densities that increase with increasing halo mass and asymptote to CDM-like profiles at large radii.  For scalar field masses $m \sim 10^{-22}$ eV, consistent with large-scale structure observations, $M_{\rm h} \sim 10^{10} \,M_\odot$ haloes have lower core densities than their Cold Dark Matter (CDM) counterparts and this alleviates the Too Big to Fail problem (TBTF) in a regime where feedback is less effective. However, higher-mass SFDM haloes with $M_{\rm h} \sim 10^{11} \,M_\odot$ are denser than their CDM counterparts at small, observationally relevant radii.  
We use rotation curves of $V \sim 100$ km s$^{-1}$ galaxies from the SPARC database to show that SFDM exacerbates the cusp/core and central density problems seen in CDM at this scale.  We conclude that if the conventional cosmological SFDM scaling relations are correct, then baryonic feedback is required to lower densities in SFDM haloes even more so than in CDM.  This motivates cosmological and self-consistent hydrodynamic simulations of SFDM to determine whether central soliton structure can be altered by realistic feedback implementations.
\end{abstract}

\section{Introduction}\label{sec:intro}

The nature of dark matter is one of the greatest puzzles in astrophysics and cosmology. 
The standard  Cold Dark Matter (CDM) model assumes that most of the matter content of the Universe is in the form of a non-interacting and non-relativistic matter component. 
Under these assumptions, the model has successfully described large-scale cosmological observations \citep{planck15,guo16,vogel14a}, but it has mismatches with observations at much smaller scales \citep{bullock17}.  Well-known issues include the cusp/core and central density problems inferred from inner rotation curve shapes \citep{flores1994,moore1994,karukes15}, the Too Big to Fail problem associated with lower-than-expected central densities of small dwarf galaxies \citep{boylan11}, and the missing low-mass galaxy problem  \citep{klypin1999,moore1999}.  

It is possible that a better understanding of baryonic processes will resolve these issues.  Supernova feedback, for example, can reduce the dark matter density in the cores of galaxies \citep{navarro96,governato12,brooks13,chan15,onorbe15,read16}, but only if the galaxy produces enough stars \citep{penarrubia12,SGK13,dicintio14,onorbe15,fitts17}.  Stellar feedback in many simulations can only efficiently remove dark matter from within a galaxy's half-light radius \citep{fitts17}.  Therefore, dark matter density discrepancies in the smallest dwarf galaxies and at large radii are more difficult to explain with baryonic feedback \citep{papa15,bullock17}. 

Alternatively, the small-scale issues may point to something deeper about the nature of dark matter.  For example, if the dark matter particles have strong self-interactions \citep{spergel00,manoj16}, the resulting flattening of density cusps into cores can alleviate  central density problems \citep{vogel12,rocha13,zavala13,oliver15,vogel16,robles17}. If the dark matter is a thermal relic of appropriate mass ($m \sim$ keV), it behaves as warm dark matter (WDM), streaming freely in the early universe to suppress small-scale power.  This completely prevents the formation of small dark matter haloes and makes dwarf-size dark matter haloes form later and with lower overall densities compared to CDM \citep{bond1982,bode2001,schneider12,ale16,horiuchi16,bozek16}.

\begin{figure}
\includegraphics[scale=.65,keepaspectratio=true]{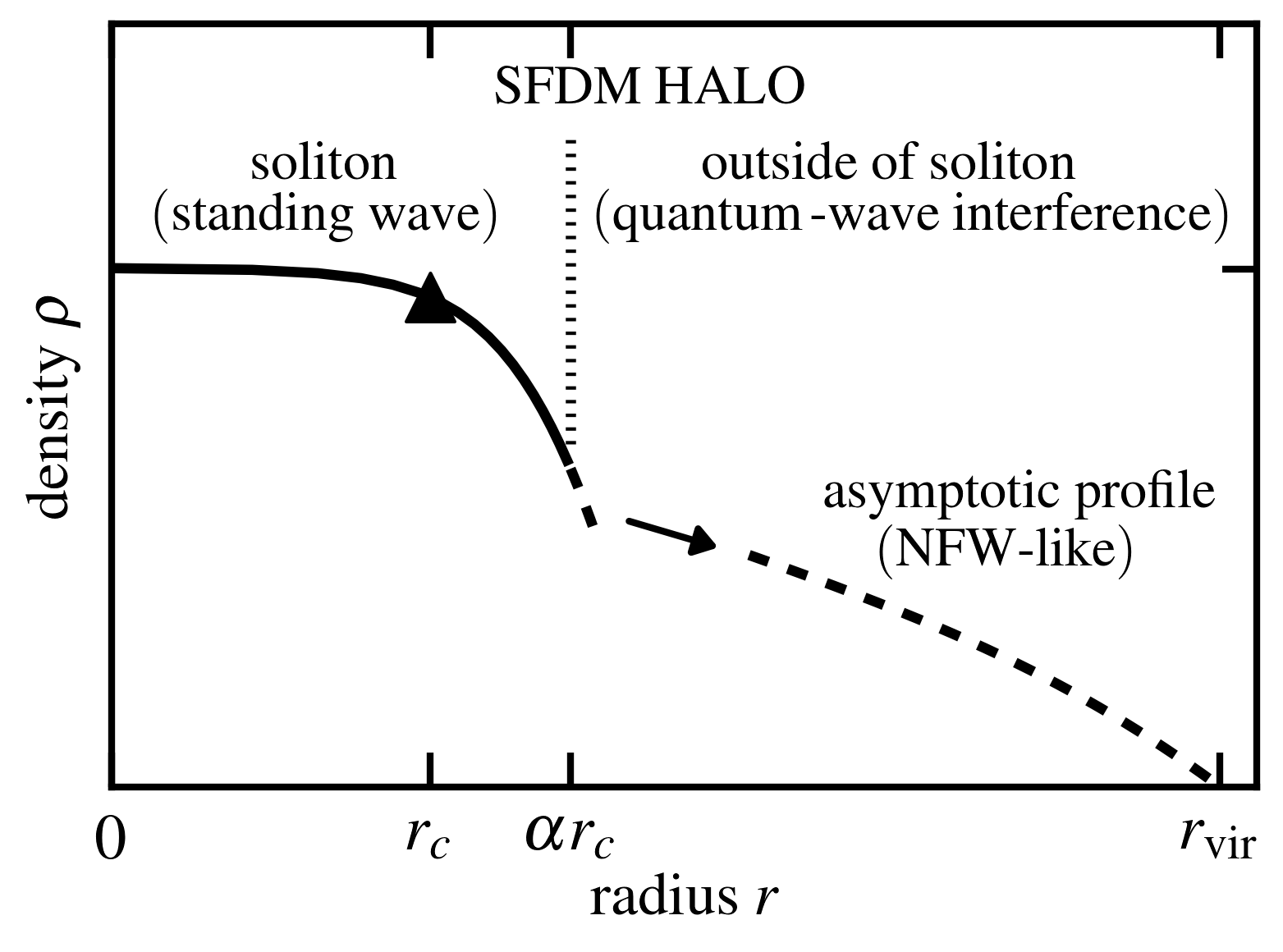} 
 \caption{Schematic diagram of the density profile of a SFDM halo.  The central region is dominated by a soliton of characteristic radius $r_c$ set by the halo virial mass $M_{\rm h}$ and scalar field mass $m$: $r_c \sim \lambda_{\rm dB} \propto m^{-1} \, M_{\rm h}^{-1/3}$.   The outer region asymptotes to a CDM-like NFW profile.  The transition from the soliton to the outer halo occurs at a radius $r = \alpha \, r_c \sim 3 \, r_c$ (vertical dotted line) and is marked by an abrupt change in average density profile slope.  In this region,  the dark matter is mildly turbulent with density fluctuations of characteristic physical size $\sim \lambda_{\rm dB}$.}
 \label{fig00}
\end{figure}

This work focuses on another possibility: that the dark matter is an ultra-light  ($m \sim 10^{-22}~{\rm eV}$) scalar field with negligible self interactions \citep{lee96,hu00,matos00,amendola06,lundgren10,hui17,suarez17}.  At early times, the field oscillates like a classical axion and has energy density that redshifts like matter  ($\rho \propto a^{-3}$).  On larges scales, the model mirrors CDM.  On small scales, however, the phenomenology is quite different.   The ultra-light mass
gives a de Broglie wavelength that is astrophysically significant ($\lambda_{\rm dB}\sim 0.1-1~{\rm kpc}$).   This produces an effective quantum pressure that suppresses power below a Jeans scale and prevents the formation of very small haloes ($M_{\rm min} \sim 10^8$ M$_\odot$).  For the larger haloes that do form, the quantum pressure prevents the formation of a central cusp that is characteristic of CDM and WDM haloes.  Galaxy haloes composed of ultra-light scalars have density profiles that are distinctive, with a central ``soliton" embedded within a familiar CDM-like outer halo \citep{matos04,harko07,sikivie09,sua11,chavanis12,rob12,rob13,bohua14,marsh14}.   
Though dark matter of this class has been explored extensively for almost two decades, it has somewhat confusingly been referred to by many different names.   Frequent terms include Bose-Einstein Condensate Dark Matter,  Ultra-light Dark Matter, Wave Dark Matter, $\psi$DM, and Fuzzy Dark Matter.  We adopt ``Scalar Field Dark Matter" (SFDM) here because a literature search showed this name to be the most common. 

 Current constraints on SFDM from structure formation and CMB data provide  lower bounds on the scalar field mass \citep{matos09,hlozek15,bozek15,sarkar16}. For example, \citet{amendola06} used Lyman $\alpha$ forest constraints to limit $m > 0.5 \times 10^{-22}~{\rm eV}$.  \citet{bozek15} and \citet{schive16} both find $m \gtrsim 1 \times 10^{-22}~{\rm eV}$ from reionization and high-redshift UV-luminosity function comparisons.  We adopt $m > 0.8 \times 10^{-22}$eV as a conservative lower limit on the scalar field mass in what follows. 
 
 Galaxy rotation curves and kinematic data for dwarf galaxies in the Local Group suggest that scalar field masses in the range $m \simeq 0.1- 5 \times 10^{-22}$eV provide better agreement on dwarf-galaxy scales than CDM.  Specifically, this mass range produces constant-density soliton cores of size $\sim 0.1-1$ \kpc \, in dwarf galaxies \citep{lora12,lora14,martinez15,robles15,lora15,chen17,calabrese16,ure17}.   Taken together with the lower limits provided by structure formation, these efforts pinpoint a mass range $m \simeq 0.8 - 5 \times 10^{-22}$ as astrophysically interesting for SFDM. While substantially higher values of $m$ are likely to be consistent with available structure formation and galaxy-scale constraints, such models do not result in observationally-relevant dark matter cores (in the absence of baryonic physics), so we do not consider such models here.

Several authors have provided analytic self-gravitating solutions for SFDM haloes in spherically symmetric configurations \citep{gleiser88,seidel94,bal98,urena02,guz04,guz06}. It is now well established that SFDM admits stable, minimum-energy configurations that are attractor solutions in the presence of small perturbations.  These stationary solutions are referred as ``solitons" \citep{sei90,seidel94,lee89,gleiser89,guz04,chava11,chava16}.

\begin{figure*}
\centering 
\includegraphics[scale=.56,keepaspectratio=true]{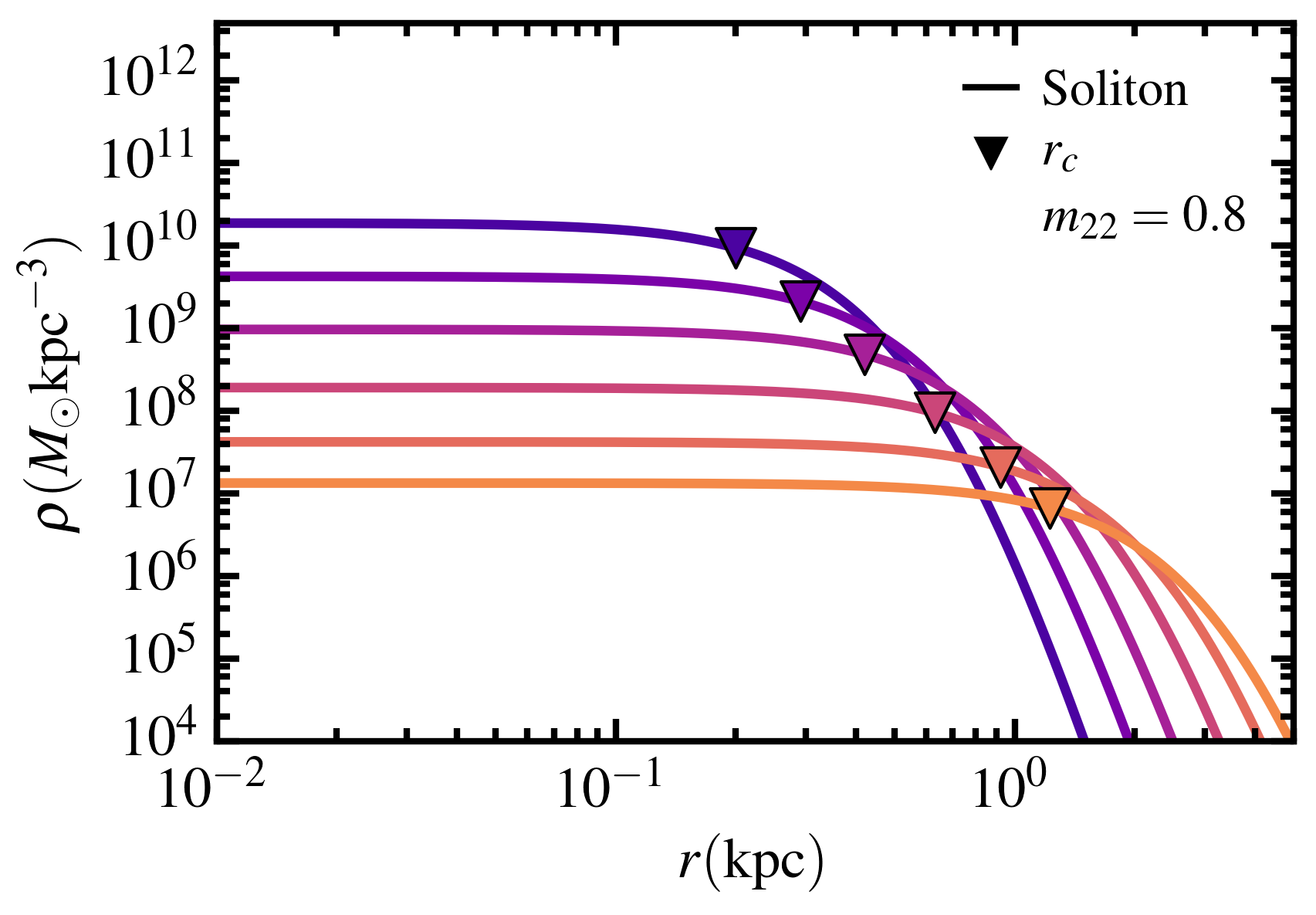} 
\includegraphics[scale=.56,keepaspectratio=true]{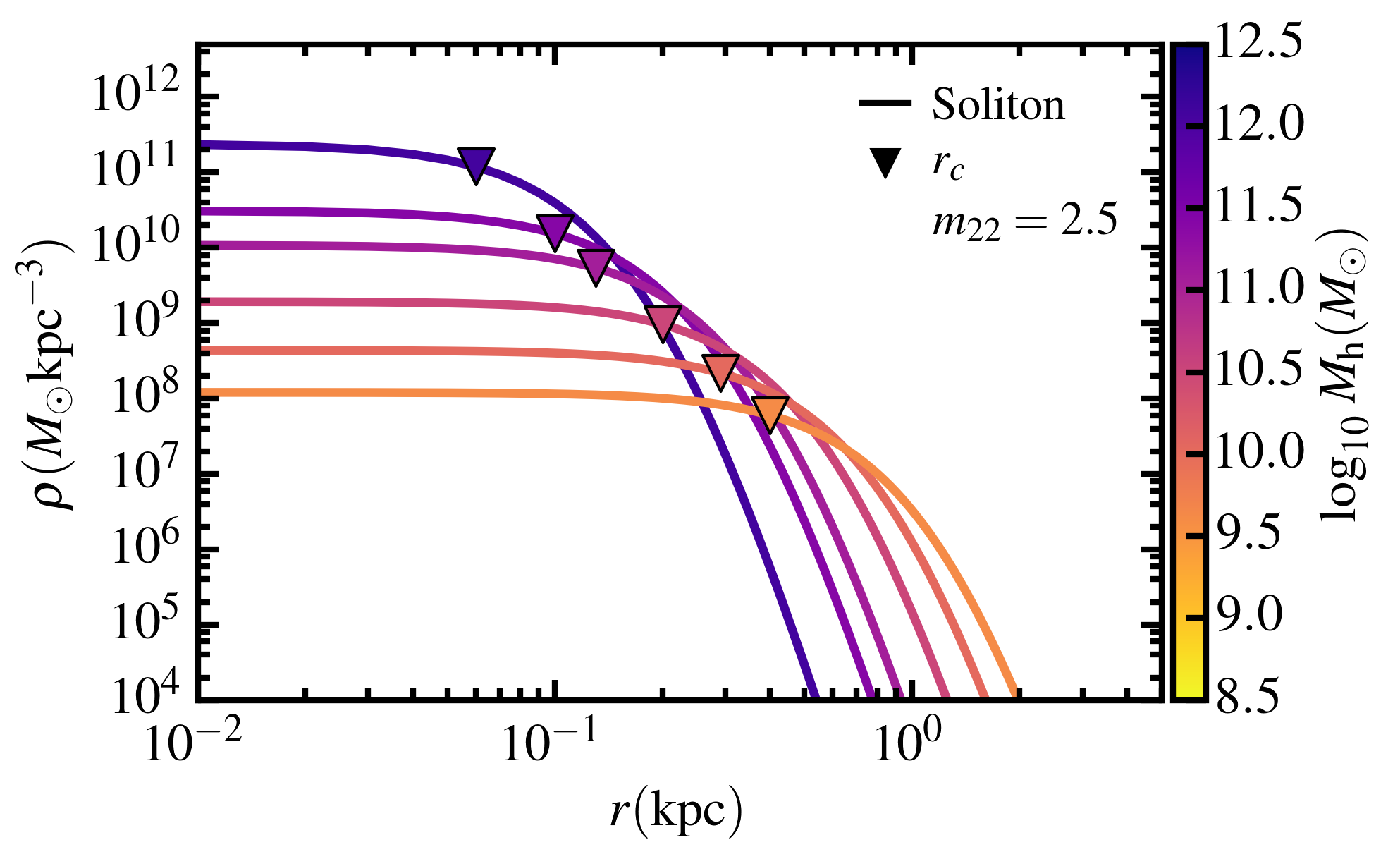} 
 \caption{Density profiles for the soliton cores in SFDM haloes according to Equations \ref{rcmh} and \ref{rhoc}.  Results are shown for scalar field masses of $m = 8 \times 10^{-23}$ eV (left) and $m = 5 \times 10^{-22}$ eV (right). The core radius for each soliton is marked as a triangle.  The color code is matched to halo virial mass (see colorbar at right).  For a fixed scalar field particle mass $m$, lower-mass haloes have larger soliton cores with lower overall density.  At fixed halo mass, the solitons are less dense and have larger radii for lighter bosons. Note that the outer dark matter envelope that is expected to surround each soliton is not shown in this figure.}
 \label{fig0}
\end{figure*}

Only recently has it been possible to conduct a fully self consistent SFDM simulation within a cosmological volume \citep{sch14}.  These DM-only SFDM simulations confirmed the existence of a compact and stable self-gravitating soliton at the centers of SFDM haloes. Further, these authors found that the central soliton is surrounded by a turbulent medium dominated by less dense fluctuations with characteristic sizes similar to that of the soliton.   Though these simulations represent a major achievement,  the demanding constraints on spatial resolution have demanded fairly small comoving volumes, and thus an exploration of the statistical halo properties was not feasible.

Non-cosmological (idealized) simulations are much less computationally expensive to run and this approach has provided a useful avenue for insight into the process of relaxation and halo collapse in SFDM.    \citet{schwabe16}, for instance, studied two-soliton merger interactions for different halo parameters. Similarly, \citet{mocz17} conducted simulations of multiple soliton cores merging, characterizing properties of the central dense soliton and the outer turbulent density field.  These studies have provided results consistent with those of \citet{sch14} and showed that the SFDM halo profile beyond the core soliton resembles a \citet[][hereafter NFW]{nfw97} profile as seen in CDM simulations.  \citet{mocz17} showed that in this outer NFW-like region, SFDM halo structure is governed by an equipartition between potential, classical kinetic, and quantum gradient energies, whereas the quantum gradient energy supports the profile inside the soliton.

 \cite{mocz17} also found that there is a dominant mode (wavelength) that contains most of the energy in the turbulent medium. Interestingly, this characteristic wavelength that seeds most of the interference sets a preferential length scale, which is observed to be about the soliton diameter.  The existence of a preferential scale for interference leads to the possibility that the average density field will not be completely smooth, but have small amplitude oscillations resulting from the net effect, the degree of smoothness is related to the size of the soliton, which itself is modified by the total mass distribution. In this work, we will not consider the modeling of the fluctuating field. However, in the Appendix we provide a way to extend our halo model to include oscillations in the density field analytically. We show that small amplitude oscillations in the density appear also as oscillations in the circular velocity profiles, which might be an observable feature in spiral galaxies.

In what follows, we provide a comprehensive method for predicting the density profiles of non-self-interacting SFDM haloes as a function of halo virial mass.  The resultant density profiles capture the expected mass distributions of a broad range of SFDM haloes and can be used to compare to observations.  Building on the results of cosmological simulations in both SFDM and CDM, our procedure matches the inner soliton prediction to the outer NFW-like profile at large radius.

In Section 2, we summarize the current understanding of inner soliton structure as a function of halo virial mass, present our formalism to connect the soliton region to the outer CDM-like profiles, and discuss the expected scatter in outer dark matter mass distributions for a given halo mass using our model. In Section 3, we use our results to confront the TBTF problem for low-mass dwarf galaxies as well as the rotation curve shapes of higher-mass galaxies and show that a single scalar field mass has difficulty resolving both problems simultaneously. Our results set a first step to model haloes that are currently beyond the SFDM simulation capabilities, motivating further progress in large-scale cosmological BEC/SFDM simulations to better constrain the scatter we predict in dark matter density profiles for a fixed halo mass and SFDM particle mass $m$. In the appendix we provide a method to extend our density profile to include the (average) oscillations of the density field. 
\section{SFDM density profiles}

Figure \ref{fig00} shows a schematic representation of our current understanding of the expected structure of an SFDM halo with virial radius $r_{\rm vir}$.  The structure is characterized by an inner soliton of radius $r_c \sim \lambda_{\rm dB}$ and an outer profile that asymptotes to the expectation for CDM haloes at large $r$.  The soliton is embedded within a turbulent medium, and envelope of dark matter that asymptotes to the CDM solution as $r \rightarrow r_{\rm vir}$.  The transition from the soliton core to the outer halo occurs at a radius $r = \alpha r_c$ where $\alpha \sim 3$ (vertical dotted line) and is usually marked by a discontinuous slope.

Below, we provide a self-consistent parametrization of the full density profile of SFDM haloes and use it to estimate the expected scatter in the dark matter mass distribution beyond the soliton radius. In our approach, we join the inner soliton to the outer region at $r = \alpha \, r_c$ and explore a range of $\alpha$ values informed by numerical simulations and physically-motivated limits.  Our values range from $\alpha \sim 2$ for dwarf-size haloes to $\alpha \sim 3-4$ for Milky Way size haloes.   We start by summarizing what is known about soliton cores and their structure as a function of halo mass.

\begin{figure*}
\centering 
\includegraphics[scale=.6,keepaspectratio=true]{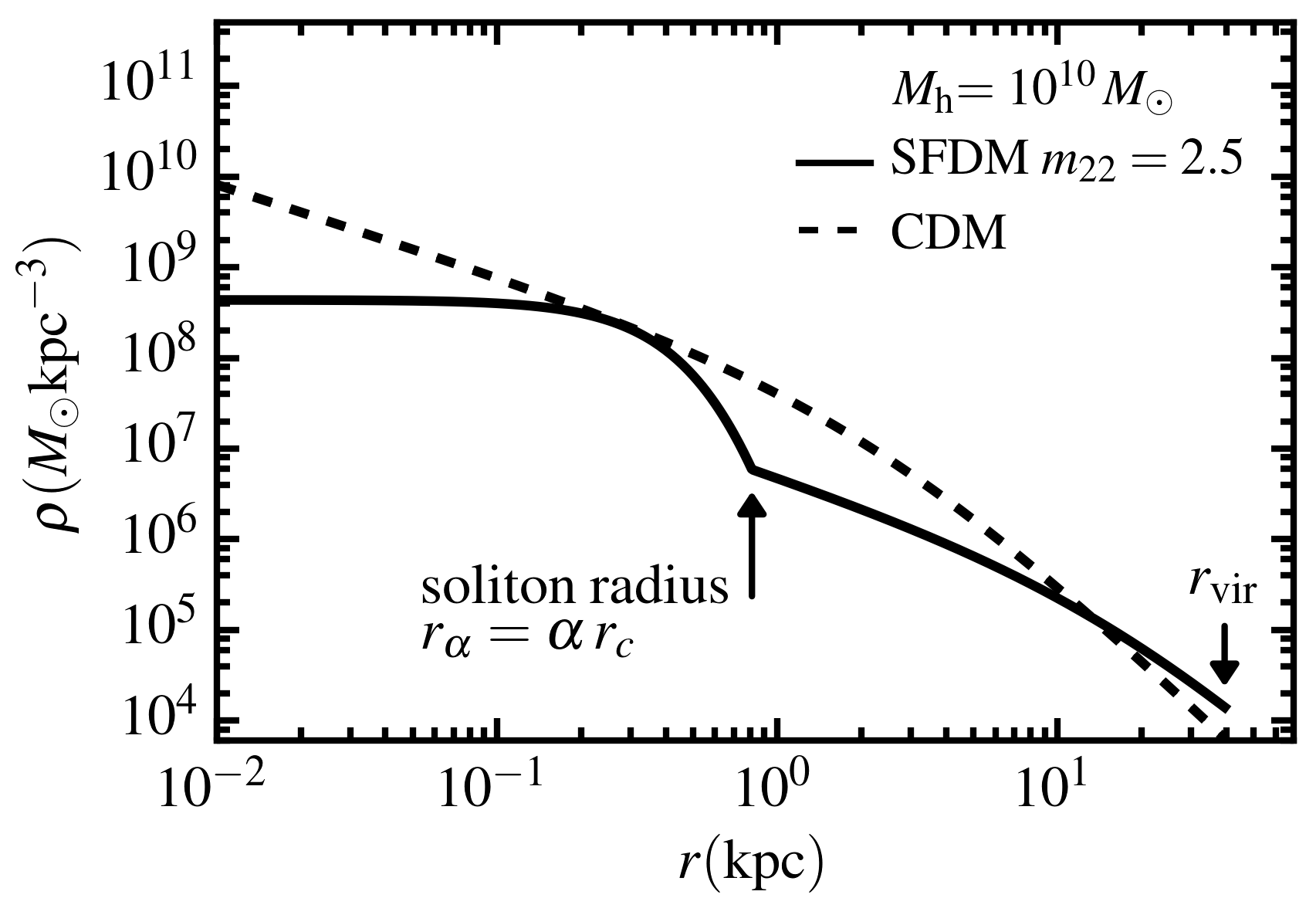} 
\includegraphics[scale=.6,keepaspectratio=true]{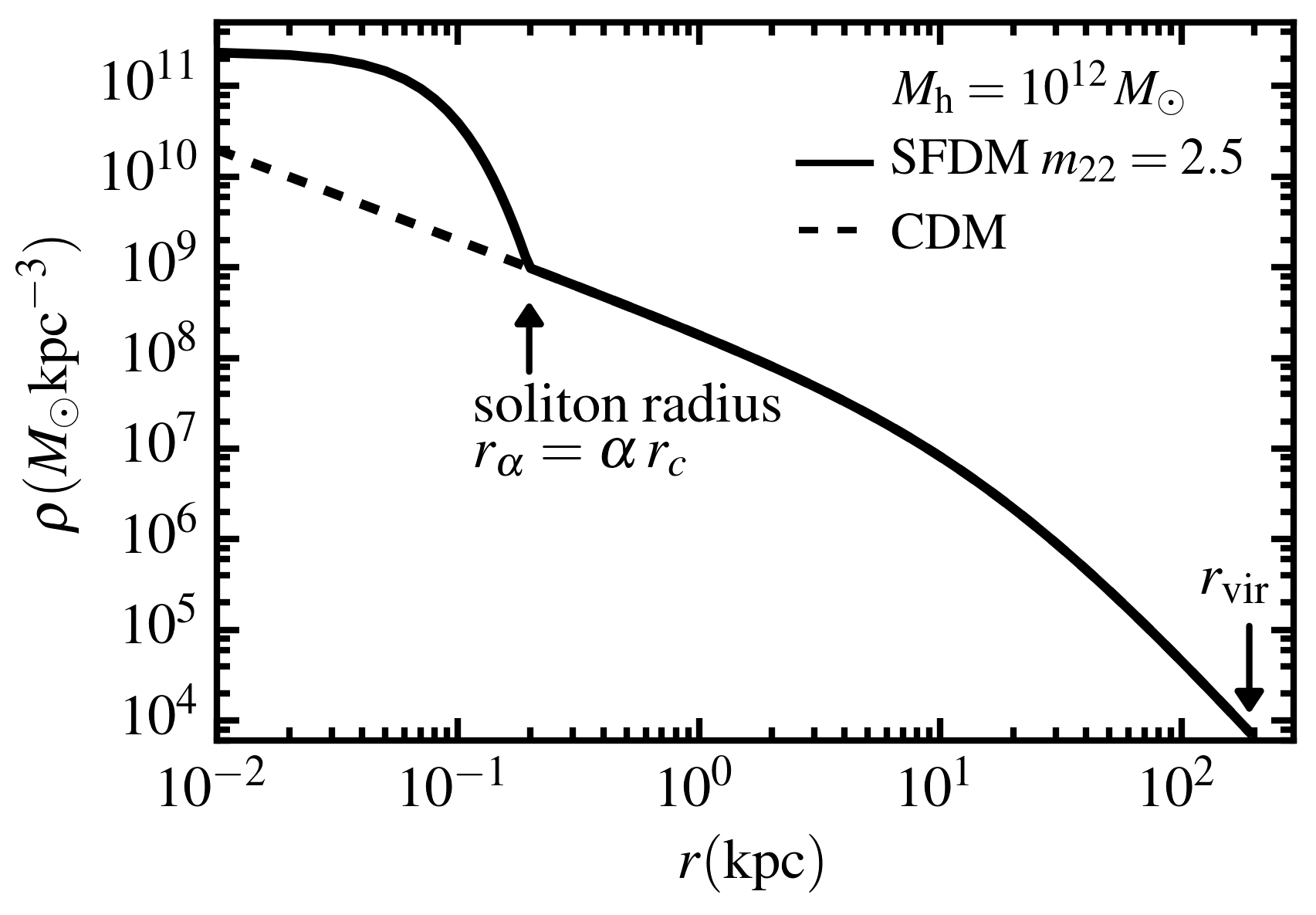} 
 \caption{SFDM (solid, with $m_{22} = 2.5$) and CDM (dashed) profiles shown for a a dwarf-size halo ($M_{\rm h} = 10^{10} M_\odot$, left) and a Milky-Way size halo ($M_{\rm h} = 10^{12} M_\odot$, right).
The SFDM halo transitions from soliton to NFW at $r_{\alpha} \equiv \alpha \,r_c$ with $\alpha = 2$  and $3$ on the left and right, respectively (see text for details).  In both cases, the soliton transition is sharp, in agreement with simulations \citep{sch14,mocz17}. The soliton core is large and of lower density than the CDM case in the dwarf halo.  In the Milky Way halo, the soliton is denser than the CDM halo at radii $r \la r_c$.}
 \label{example}
\end{figure*}

\subsection{Soliton Cores}
\label{sec:solitons}

The structure of a SFDM halo is governed by the Schrodinger-Poisson (SP) equations for a self-gravitating scalar field with associated gravitational potential V:
\begin{eqnarray}
i\hbar \frac{\partial \psi}{\partial t} &=& -\frac{\hbar^2}{2m}\nabla^2\psi + m V\psi \nonumber \\
\nabla^2 V &=& 4\pi G(\rho-\overline{\rho}),
\label{sp}
\end{eqnarray}
where $\psi$ is the wave-function and $\rho=\lvert \psi\rvert^2$ is the DM density \citep[see, e.g.][]{chava11}.

Using cosmological simulations,  \citet{sch14} found that all collapsed haloes develop a central soliton core described by the ground state solution of the SP equations.  They found that the central core region was well fit by the profile
\begin{equation}
\rho_{\rm sol}(r)=\frac{\rho_c}{\left(1+0.091\,\left(\frac{r}{r_c}\right)^2\right)^8},
\label{sol}
\end{equation}
where $\rho_c$ is the central density given by
\begin{equation}
\rho_c=1.93\times 10^7\,m_{22}^{-2}\left(\frac{r_{\rm c}}{1 \,\kpc}\right)^{-4} M_{\odot} \kpc^{-3}, 
\label{rho}
\end{equation}
with
\begin{equation}
 m_{22} \equiv \frac{m}{10^{-22}\,{\rm eV}/c^2}.
 \end{equation}
Note that for fixed $m$, the soliton density scales inversely with soliton size such that the smallest solitons are physically denser: $\rho_c \propto m^{-2} \, r_c^{-4}$.  These scalings between density, radius, and scalar field mass are demanded by the symmetry of the SP equations \citep{schive14prl}.

The soliton structure is governed by the global potential envelope of the halo it inhabits.  We expect its size to be similar to the de Broglie wavelength $r_c \sim \lambda_{\rm dB} \equiv h/(m \, v)$.  In a collapsed halo of virial mass $M_{\rm h}$, the characteristic velocity increases with halo mass as $v \propto M_{\rm h}^{1/3}$, which implies $r_c  \propto m^{-1} \, M_{\rm h}^{-1/3}$.  Indeed,  \citet{sch14} noted that the soliton cores in  their cosmological simulations obeyed precisely this expected scaling.  At redshift zero they found
\begin{equation}
r_{c}=1.6 \, {\rm kpc} \, \bigg (\frac{M_{\rm h}}{10^9 M_{\odot}} \bigg)^{-1/3} \,  m_{22}^{-1} ,
\label{rcmh}
\end{equation}
where M$_{\rm h}$ uses the \citet{bryan98} definition of virial mass.  Together, equations \ref{rho} and \ref{rcmh} imply that 
\begin{equation}
\rho_c = 2.94 \times 10^6  \, M_{\odot} \kpc^{-3} \bigg (\frac{M_{\rm h}}{10^9 M_{\odot}} \bigg)^{4/3}  \, m_{22}^{2} .
\label{rhoc}
\end{equation}
If we define the soliton core mass as $M_{\rm c}\equiv 4\pi\rho_{\rm c}\,r_{\rm c}^3/3$ then we have
\begin{equation}
M_{\rm c} = 5.04 \times 10^{7} \, M_\odot \, \bigg (\frac{M_{\rm h}}{10^9 M_{\odot}} \bigg)^{1/3} \,  m_{22}^{-1} \, .
\label{mc}
\end{equation}
One implication of these relations is that the smallest haloes will have the largest soliton core radii and lowest density soliton cores.   Also, as halo mass increases, the soliton core size as a fraction of virial radius becomes insignificant $r_c/r_{\rm vir} \propto$ M$_{\rm h}^{-2/3}$:
\begin{equation}
\frac{r_{\rm c}}{r_{\rm vir}}=6.20\times 10^{-2}\,\left(\frac{M_{\rm h}}{10^9\,M_{\odot}}\right)^{-2/3}\,m_{22}^{-1}\,.
\label{frac_rc}
\end{equation}
Similarly, the fraction of a halo's mass locked up in the soliton core 
decreases rapidly with halo mass $M_c/M_{\rm h} \propto M_{\rm h}^{-2/3}$:
\begin{equation}
\frac{M_{\rm c}}{M_{\rm h}}=5.04\times 10^{-2}\,\left(\frac{M_{\rm h}}{10^9\,M_{\odot}}\right)^{-2/3}\,m_{22}^{-1}\,.
\label{frac_mc}
\end{equation}
 The ratio of the circular velocities at the core and virial radius does not depend on either $M_{\rm h}$ or $m$: $V_{\rm c}(r_{\rm c})/ V_{\rm c}(r_{\rm vir}) \sim 0.9$.

Finally, we note that the product $r_{\rm c}\,M_{\rm c}$ is independent of halo mass:
\begin{equation}
r_{\rm c}\,M_{\rm c}=8.06 \times 10^7 \,m_{22}^{-2}\,M_{\odot}\,\kpc \,.
\label{rc_mc}
\end{equation}
Eq. ~\ref{rc_mc} means that the specific angular momentum of an object on a circular orbit within the soliton core depends only on the scalar field mass, not on the halo mass itself: $j_c = \sqrt{G \, r_c \, M_c}  = 18.6$ $m_{22}^{-1}$ kpc km s$^{-1}$.  This is very similar to the characteristic angular momentum scale of the quantum field, $\hbar/m =19 \, m_{22}^{-1}$ kpc km s$^{-1}$.

Figure \ref{fig0} uses Equations \ref{sol}, \ref{rcmh} and \ref{rhoc} to show the central soliton profiles for SFDM haloes over a range of virial masses (color bar). The left and right panels assume scalar field masses of $m_{22}=0.8$ and $m_{22}=2.5$, respectively.   Note that for $m_{22}=0.8 \, (2.5)$, haloes smaller than $\rm M_{\rm h}\sim 10^{9.5} \, (10^{8.5}) \, M_{\odot}$ are unstable due to the quantum pressure and may not form, therefore haloes below these masses are not plotted.

We note that although the precise factors in Eqs. \ref{rcmh} and \ref{rhoc} are empirical, being best-fit values for the cosmological simulations of \citet{sch14}; we will assume these values for the rest of our derivation. A study of how these values vary in different numerical implementations is out of the scope of this work.

\subsection{The Outer Envelope}
\label{sec:envelopes}

We aim to define a complete SFDM density profile given a particle mass $m_{22}$ and a halo mass $M_{\rm h}$. 
 \citet{sch14} found that the soliton profile in Equation \ref{sol} was only a good fit for $r \le \alpha \, r_c \simeq 3 \,r_c$.  Beyond this radius, the solitonic cores were seen to transition to a CDM-like profile similar to the \citep{nfw97} form 
\begin{equation}
\rho_{\rm NFW}(r)=\frac{\rho_s}{(r/r_s)(1+r/r_s)^2},
\label{nfw}
\end{equation}
where $r_s$ is the radius where the logarithmic slope of the density profile is $-2$.

A natural way to model an SFDM halo out to $r_{\rm vir}$ is to transition from the soliton profile (\ref{sol}) to an NFW profile (\ref{nfw}) at some specified radius $r_{\alpha}$.   Several authors have followed this approach \citep{marsh14,alma17,bernal18}; however, past choices for the transition radius were not based on simulation results but rather relied on the estimate $r_\alpha = r_c$ or finding the point where the density log-slopes are equal.  However, SFDM haloes in simulations \citep{sch14,mocz17} all show a transition radius several times larger than $r_c$ and reveal that the density slopes can differ significantly at the transition point.  Specifically, the transition from solitonic core to NFW profile is sharp, with the soliton almost superimposed on top of the NFW.

\begin{figure}
\centering 
\includegraphics[scale=.6,keepaspectratio=true]{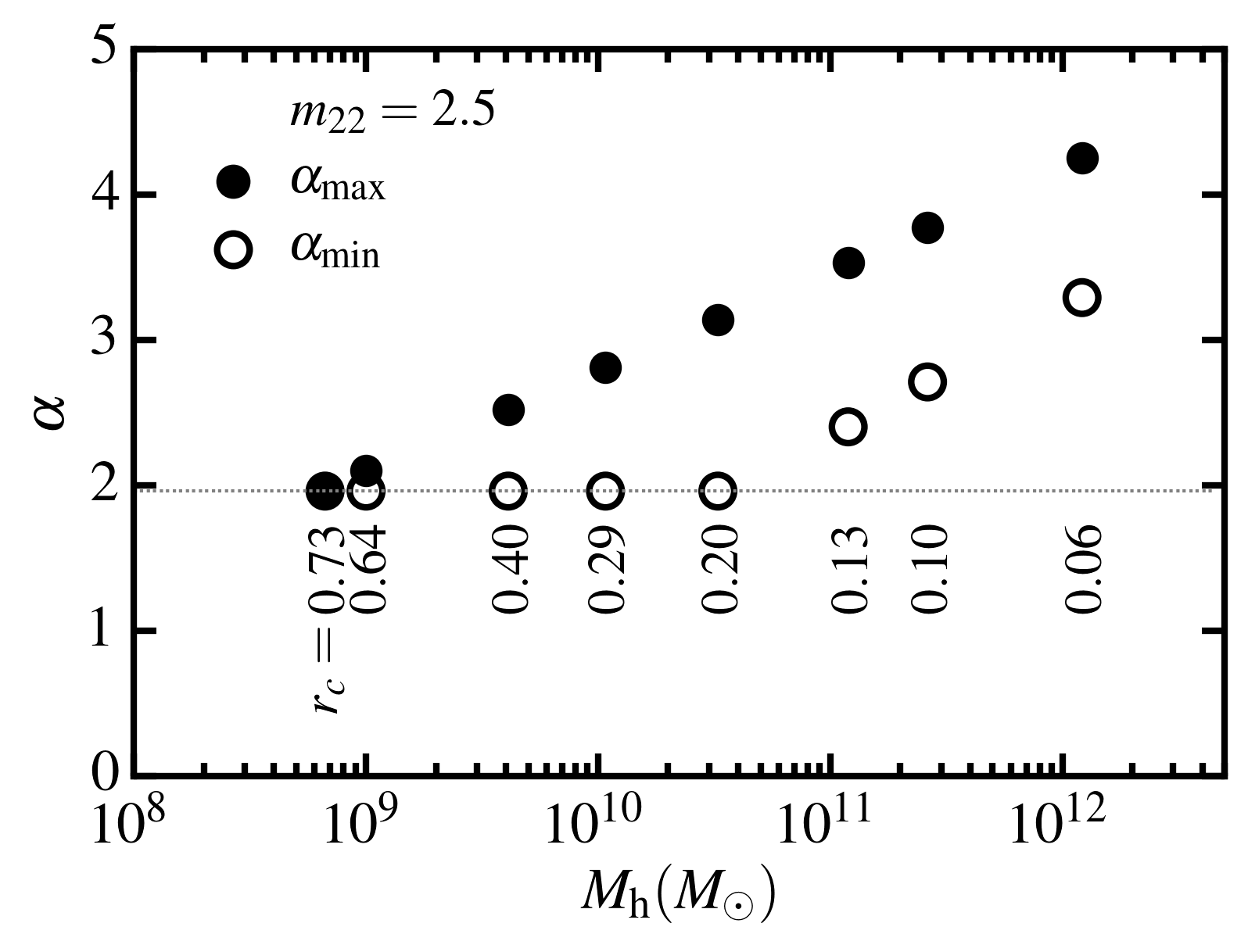} 
 \caption{Range of soliton transition radii allowed as a function of halo mass parameterized by $\alpha$, where the transition radius is $r_\alpha \equiv \alpha \, r_c$ and $r_c$ is the soliton core radius (see Equations \ref{alpha} and \ref{rhotot}).  The numbers show the value of $r_c$ in units of kpc for several halo masses.}
 \label{fig:alpha}
\end{figure}

Furthermore, SFDM simulations do not predict a unique transition radius due to the turbulence in the field \citep{sch14, schwabe16,vel16}.
The difficulty in providing a single profile for the total halo has recently led to more complex assumptions. For instance,  \citet{lin18} adopted a mode decomposition of the scalar field wave function that comes from simulations of \citet{sch14}. The authors fit the inferred distribution function from simulations to various classical particle distribution functions and found that the fermionic King model \citep{chava98} provided a good fit.  A different approach was given in \citet{bar18}, who reinterpret Eq. \ref{rcmh} as a statement that the energy per unit mass of the soliton is equal to that of the total halo and develop their analysis under this assumption.

We will follow a phenomenological 
approach to link the inner halo with the outer halo.
To capture the features observed in SFDM simulations and account for the (relatively small) variations in the transition radius for a given halo mass, we explore a range of radii that mark the transition from soliton to outer halo:
\begin{equation}
\label{alpha}
r_{\alpha} = \alpha \, r_c \, .
\end{equation}
With this choice, the total density profile is
\begin{equation}
        \rho( r)=
        \left\{ \begin{array}{ll}
            \rho_{\rm sol}(r ) & 0\leq r \leq r_{\alpha} \\ [0.5ex] 
            \rho_{\rm NFW} (r ) & r_{\alpha} \leq r \leq r_{\rm vir}. 
        \end{array} \right.
        \label{rhotot}
    \end{equation}
We now must fix two parameters for both the inner soliton piece and outer NFW piece in Equation \ref{rhotot}.   Given $m_{22}$ and $M_{\rm h}$, Equations \ref{sol}, \ref{rcmh}, and \ref{rhoc} define the $\rho_{\rm sol}(r)$ completely.  For the outer NFW piece (Equation \ref{nfw}) we need two additional constraints in order to set $\rho_s$ and $r_s$. For these we impose density continuity at $r_\alpha$
\begin{equation}
\label{xxx}
\rho_{\rm sol}(r_\alpha)  =  \rho_{\rm NFW} (r_\alpha) \, ,
\end{equation}
and mass conservation within $r_{\rm vir}$
\begin{eqnarray}
\label{yyy}
M_{\rm h}  &=& 4\pi \int^{r_{\rm vir}}_0 \rho(r')r'^2dr' \\
    &=& 4\pi \int^{r_\alpha}_0 \rho_{\rm sol}(r')r'^2 dr' + \int^{r_{\rm vir}}_{r_\alpha} \rho_{\rm NFW}(r')r'^2dr' \nonumber .
\end{eqnarray}
By assuming mass conservation, we are explicitly assuming that the quantum pressure is not sufficient to result in the existence of significant additional mass beyond the virial radius compared to CDM.  This should be a reasonable approximation for all but the smallest haloes (with halo masses just above the suppression mass for a given $m$). 
\begin{figure*}
\centering 
\includegraphics[scale=.6,keepaspectratio=true]{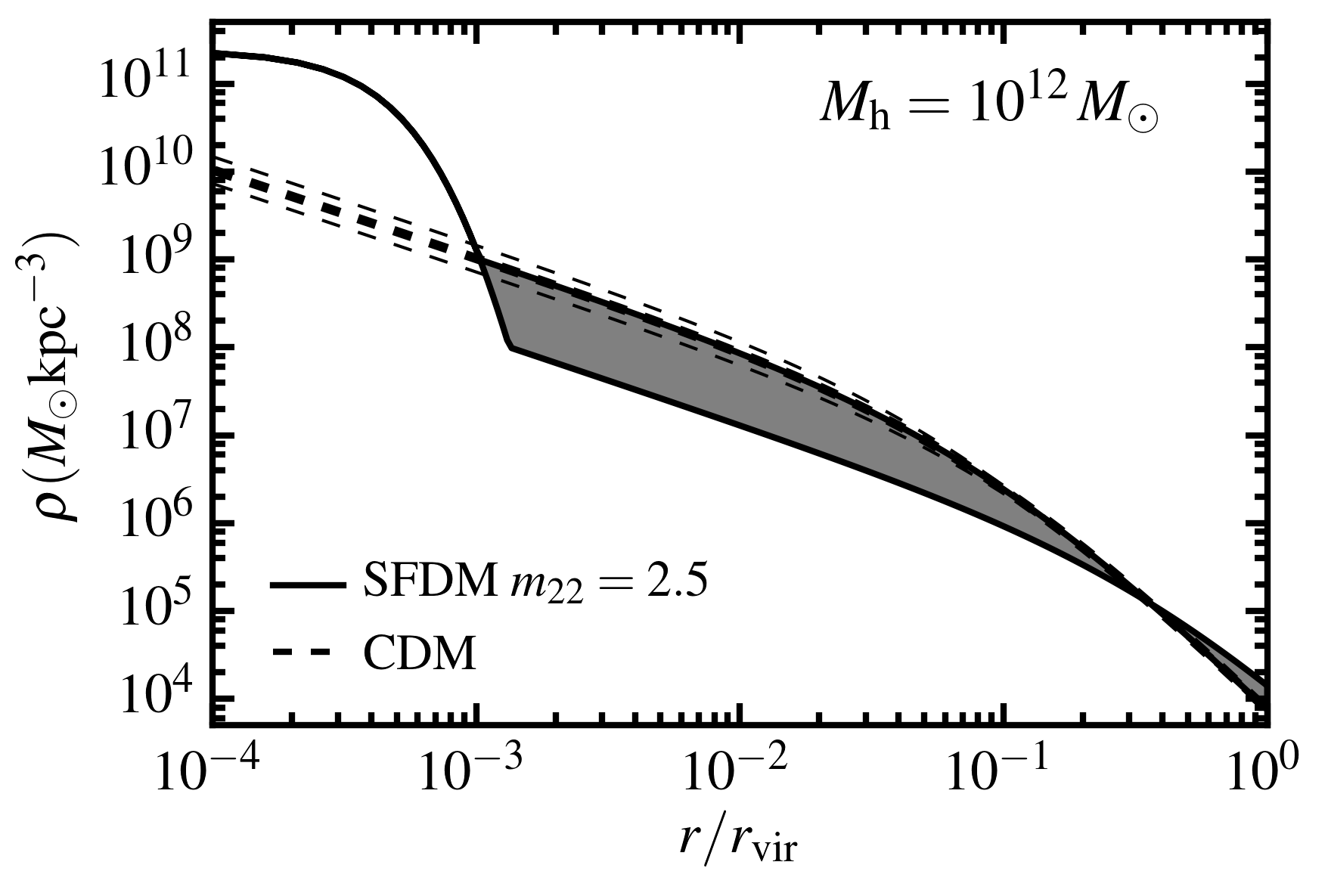} 
\includegraphics[scale=.6,keepaspectratio=true]{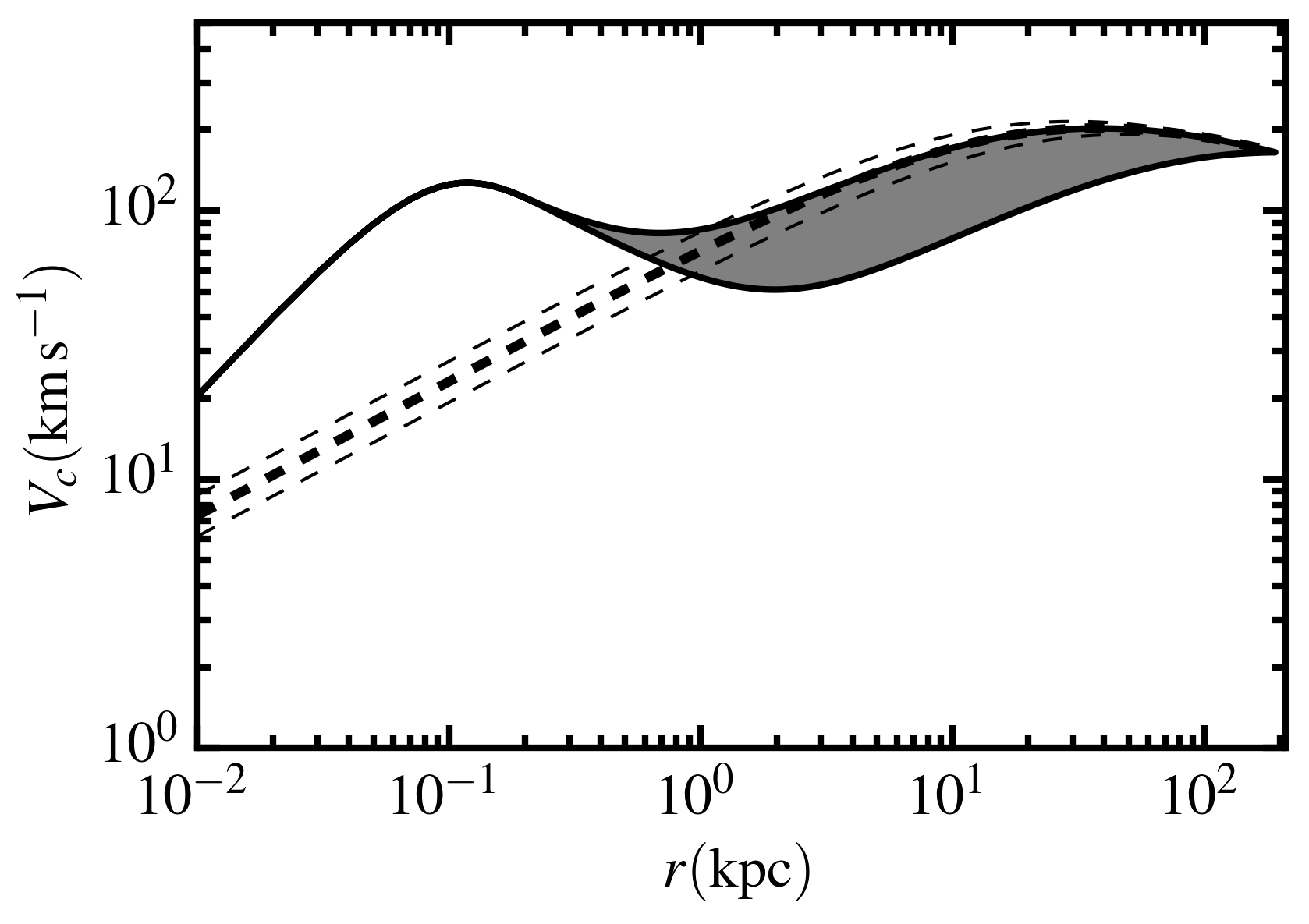} \\
\includegraphics[scale=.59,keepaspectratio=true]{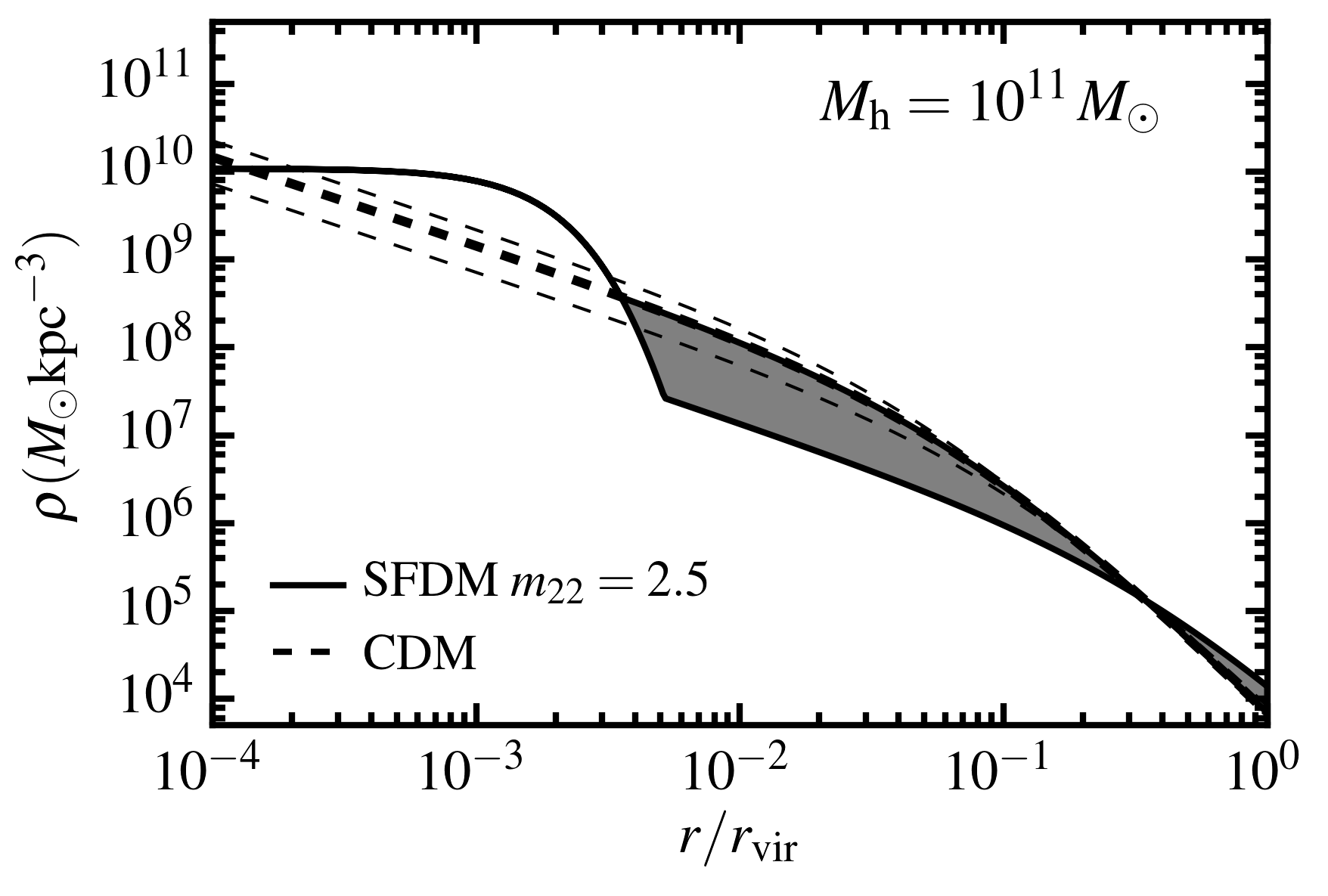} 
\includegraphics[scale=.59,keepaspectratio=true]{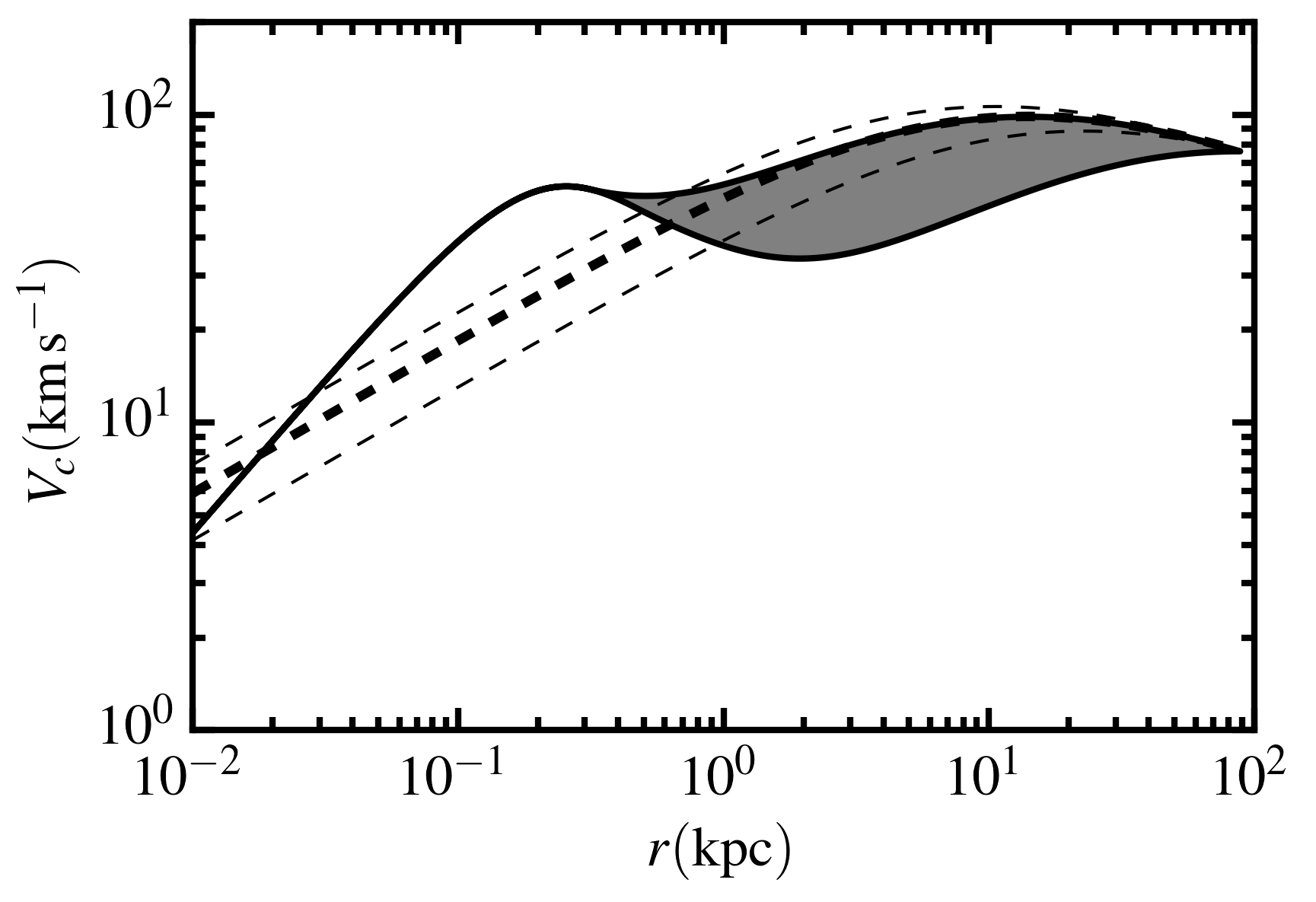} \\
\includegraphics[scale=.6,keepaspectratio=true]{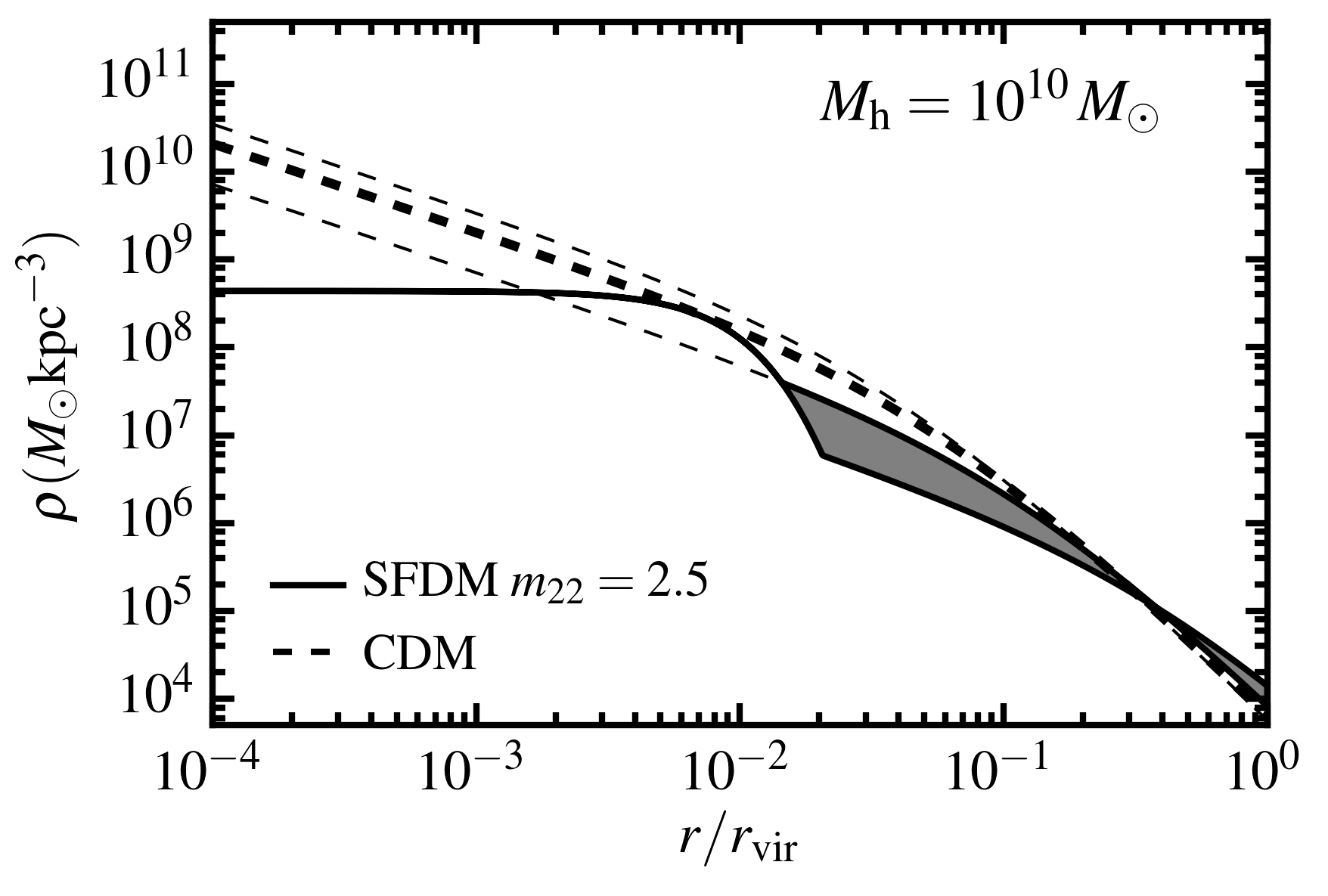} 
\includegraphics[scale=.6,keepaspectratio=true]{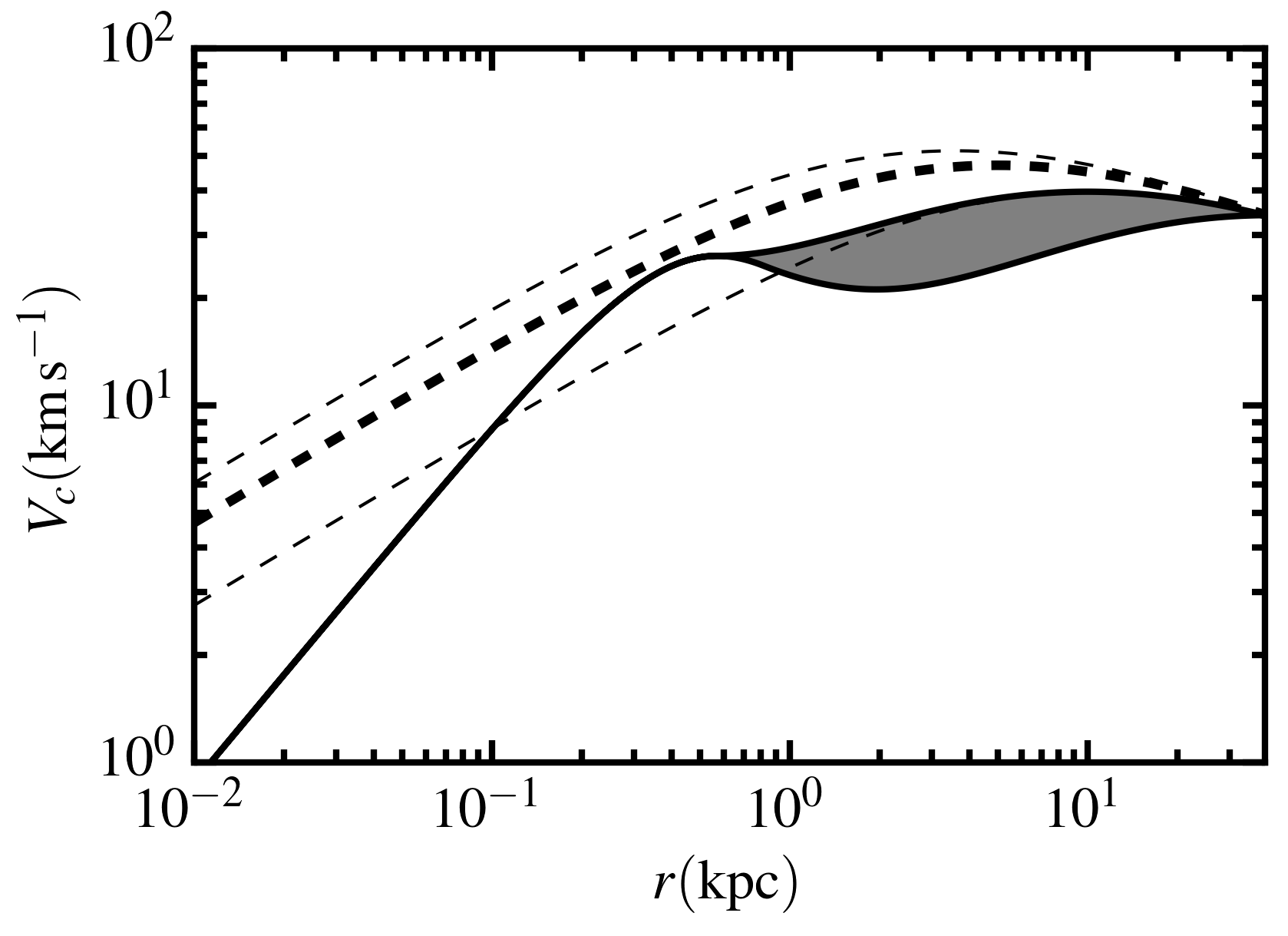} 
 \caption{Density profiles (left) and rotation curves (right) for SFDM (solid) and CDM (dashed) haloes of three example masses: 
 $M_{\rm h} = 10^{12}$ (top), $10^{11}$ (middle), and $10^{10}$ (bottom) $M_\odot$.  
 The light-dashed CDM lines show $\pm 1\sigma$ scatter in the halo concentrations at fixed halo mass \citep{mac08}.  
The density profiles are shown as a function of radius normalized by the halo virial radius in order to emphasize that the size of the soliton core is proportionally larger in 
 the lower-mass haloes.  The associated circular velocity curves (right) are plotted as a function of unnormalized radius to provide a sense of the physical scale.  
 The shaded regions in the SFDM  profiles allow for a 
 range of soliton transition radii ($r_\alpha = \alpha \,r_c$, see Figure \ref{alpha}).  In the smallest haloes, the minimum soliton size 
 ($\alpha_{\rm min}$) is set by demanding a local maximum in the circular velocity curve (lower right panel).  In the largest haloes, $\alpha_{\rm min}$ is set by demanding that the peak circular velocity within the soliton does not exceed the global circular velocity of the halo (upper right).  For all masses, the maximum soliton radius is set by requiring that the maximum circular velocity is reached within $r_{\rm vir}$. 
 }
 \label{rho_vc}
\end{figure*}

We emphasize that the outer NFW profile of the SFDM halo will not necessarily track the NFW profile for the same mass halo in CDM.   The effective concentration ($c = r_{\rm vir}/r_s$) and normalization of the two NFW haloes can be different because a non-negligible portion of the SFDM halo mass may be locked up within the central soliton.  
Of course, the global structure of CDM haloes informs what is plausible for SFDM haloes in that we generally expect the mass and potential well depth at the virial radius to be similar when $\lambda_{\rm dB} \ll r_{\rm vir}$. Fortunately, CDM halo concentrations are well understood from cosmological simulations \citep[e.g.][]{bullock01}.  In what follows we assume the \citet{planck15} cosmology and the concentration-mass relation from \citet{dutton14} to determine the CDM prediction. 

Figure \ref{example} shows example profiles for a Milky Way size halo (right) and a dwarf-size halo (left) for characteristic choices of $\alpha$ motivated below (3 and 2 respectively).  We adopt $m_{22}$=$2.5$ in this example.  Note the difference between CDM and SFDM is largest for the dwarf halo. This will generally be the case: the fractional mass contained within the soliton core increases with decreasing halo mass.  Indeed, the soliton core in the dwarf halo is less dense than the CDM cusp and has a size comparable to the half-light radius of a classical dwarf spheroidal ($\sim 500$ pc). 
Conversely, the soliton in a Milky-Way-like system at a radius of $r_c \sim 100$ pc is even denser than CDM; such a halo has a very similar density in SFDM and CDM at a typical half-light radius for a Milky Way-size galaxy ($r_{\rm gal} \sim 3$ kpc). 

Cosmological SFDM simulations find values of $\alpha \sim 3$, though these simulations exist only for haloes in a limited mass range $M_{\rm h}\sim 10^{11} \rm M_{\odot}$ \citep{sch14}.  Idealized simulations suggest that the transition occurs approximately at the radius where the potential energy density is comparable to the total quantum kinetic energy density \citep{mocz17}. The value of $\alpha$ can be modified by constructive and destructive interference of the intrinsic phase of the soliton with the uncorrelated phases of the surrounding medium. Our goal is to define a range of $\alpha$ values that are plausible at each halo mass scale. 

We bracket minimum ($\alpha_{\rm min}$) and maximum ($\alpha_{\rm max}$) soliton transition radii by imposing several simple constraints on the circular velocity profile.  These constraints are motivated by simulation results and physical plausibility.  First, we demand that the soliton produces a local maximum in the circular velocity curve $V_c(r)$ at a radius $r \sim r_c$.  Second, we ensure that the global circular velocity curve of the SFDM halo has a maximum that does not exceed its CDM counterpart (qualitatively that the potential well is never deeper than it would be in CDM).  Finally we demand that the radius where maximum circular velocity of the halo occurs $r_{\rm max}$ is within $r_{\rm vir}$.  

Figure \ref{fig:alpha} shows the range of $\alpha$ values that are allowed by our conditions as a function of halo mass.  For low-mass haloes, the condition that matters most in setting $\alpha_{\rm min}$ is the existence of a gravitationally-dominant soliton (a local maximum in $V_c(r)$).  For larger haloes ($M_{\rm h} \gtrsim 10^{11} M_\odot$), $\alpha_{\rm min}$ is set by demanding $V_{\rm max}^{\rm SFDM} \le V_{\rm max}^{\rm CDM}$. For haloes of all masses, $\alpha_{\rm max}$ is set by the condition $r_{\rm max} \le r_{\rm vir}$.   In a dwarf-size halo ($M_{\rm h} \sim 10^{10} M_\odot$), the $\alpha$ range is $\sim 2-3$.   In Milky-Way mass haloes, the $\alpha$ range is $\sim 3 - 4$.  For reference, the size of the soliton core expected at each halo mass (in kpc) is displayed below the $\alpha = 2$ line in Figure \ref{fig:alpha}.  Note that even though $\alpha$ values are slightly larger for more massive haloes, the value of $r_c$ decreases even faster, meaning that the physical extent of the soliton-dominated region ($\alpha \, r_c$) decreases with increasing halo mass.\footnote{Note that while we naively expect similar scaling to even higher masses associated with galaxy clusters, these haloes experience more recent mergers and are more dynamically influenced by large substructures.  Quantum interference is therefore likely more intricate in these systems.  Cosmological simulations at these higher mass scales will be required before we are able to make confident predictions.}

\begin{figure*}
\centering 
\includegraphics[scale=.3,keepaspectratio=true]{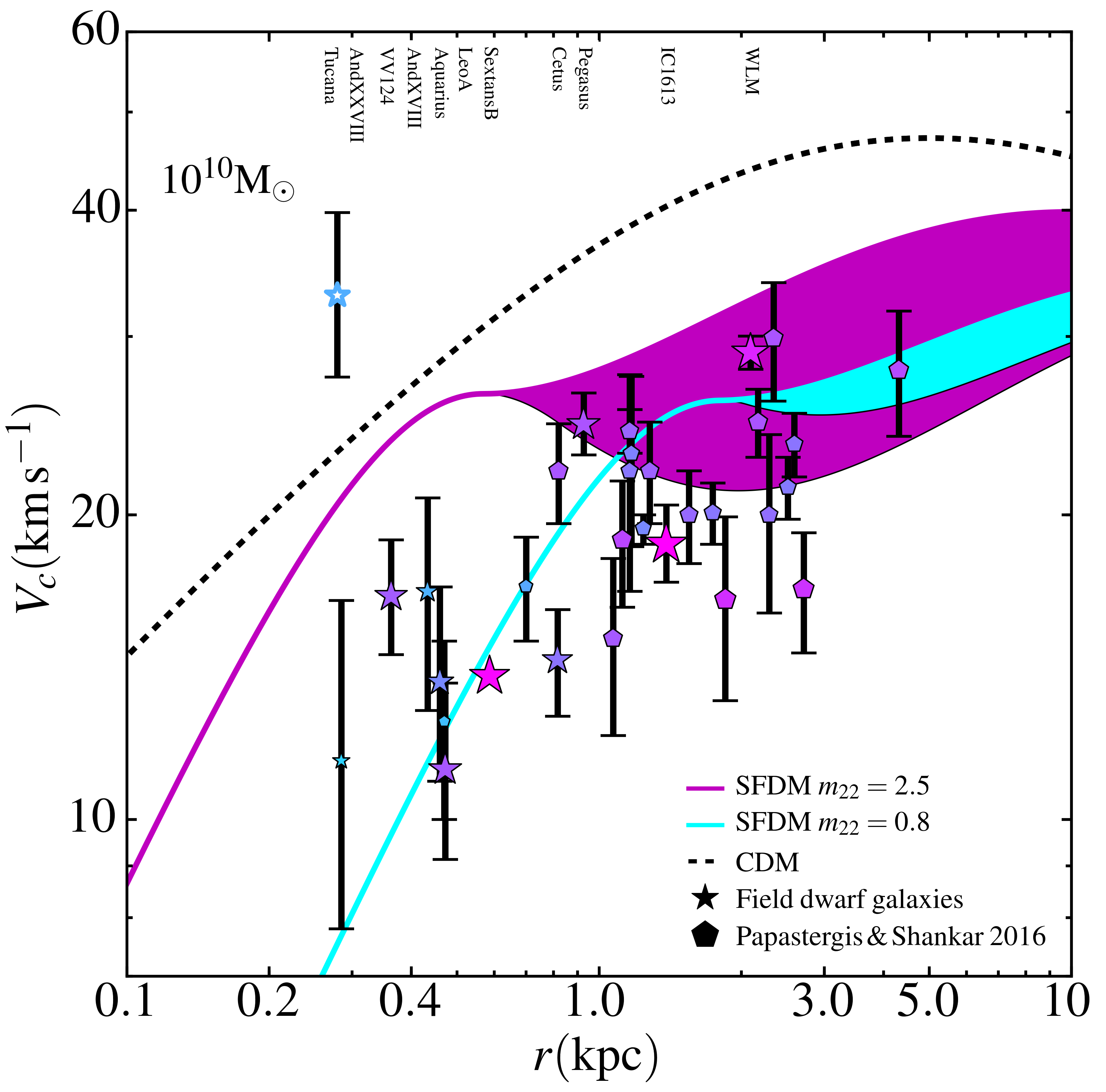}
\includegraphics[scale=.3,keepaspectratio=true]{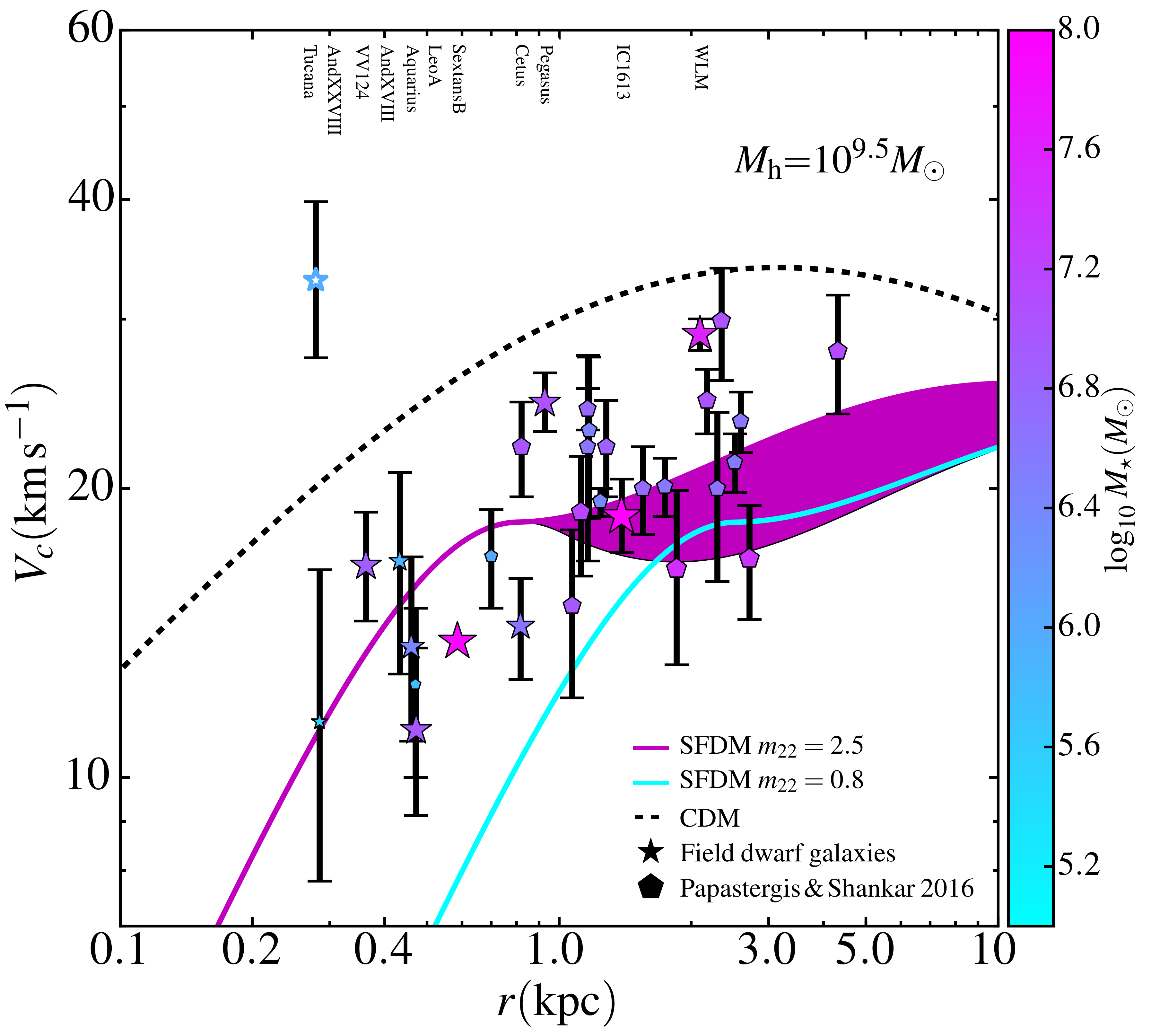}
 \caption{Too-big-to-fail halo comparison for  CDM (dashed) and SFDM (magenta for $m_{22}=2.5$ and cyan for $m_{22}=0.8$).   Stars show half-light circular velocities for non-satellite Local Group dwarf galaxies and pentagons show isolated field dwarf galaxies.  Symbol sizes and colors vary with galaxy stellar mass as indicated by the color bar.    The two panels compare CDM and SFDM predictions at fixed halo mass of $M_{\rm h}=10^{10}M_{\odot}$ (left) and $M_{\rm h} = 10^{9.5} M_\odot$ (right), which are typical of TBTF haloes.  The CDM predictions (at median concentration) are too dense relative to observations, as expected.  The SFDM haloes are a better match to the densities of observed galaxies; this is particularly notable at large radii (well beyond the galaxies' half-light radii), where baryonic feedback is less effective at producing cores in CDM haloes.}
\label{fig:tbtf}
\end{figure*}

Figure \ref{rho_vc} provides examples of how SFDM halo profiles (left) and circular velocity curves (right) differ from CDM expectations for three example halo masses: a Milky Way ($M_{\rm h} = 10^{12} M_\odot$, top), a small spiral ($M_{\rm h} = 10^{11} M_\odot$, middle), a dwarf galaxy ($M_{\rm h} = 10^{10} M_\odot$, bottom).  In the left panels, we plot the density as a function of $r/r_{\rm vir}$.  In the right panels, radii are ploted in physical (unscaled) units.  In each case, solid lines correspond to SFDM and dashed lines to CDM. The shaded regions in the SFDM haloes span $\alpha_{\rm min}$ to $\alpha_{\rm max}$.  Note that in the smallest haloes, the soliton  dominates within an appreciable ($\sim 1 \%$) fraction of the virial radius and has a lower core density than CDM.  For larger haloes, however, the soliton can be as small as $\sim 0.1 \%$ of the virial radius and its density is larger than the CDM halo in this region.  
For dwarf-size haloes, SFDM circular velocity curves are expected to lie below and rise more quickly than corresponding $V_{\rm c}(r)$ in CDM; for hosts of $V_c \sim 100 - 200$ km s$^{-1}$ galaxies, SFDM haloes are both denser and more peaked at small radii than CDM.    This unique behavior may provide a means to discriminate between the theories: it is substantially different from what is found in other alternative dark matter models, where the core region is generally less dense than the density of a CDM halo of the same virial mass \citep[see][and references therein]{robles17}.
\vspace{1cm}

\section{Comparison with galaxy data}

\subsection{The Too Big to Fail problem and SFDM haloes}

As originally cast, the TBTF problem with CDM (DM-only) simulations refers to the fact that the central densities of the most massive subhaloes in a Galaxy-size hosts are higher than the central densities observed in dwarf galaxies  of the Milky Way \citep{boylan11,boylan12} and Andromeda \citep{tol14,kirby14}.  A related problem exists for galaxies in the Local Group that are not satellites of larger hosts \citep{garrison-kimmel2014a} and for dwarf galaxies in the field \citep{papa15,papa16,ferrero12}.  The second incarnation associated with non-satellite galaxies is regarded as more problematic.  This is because satellite subhaloes are likely to end up less dense than seen in DM-only simulations owing to enhanced mass loss associated with the central galaxy potential \citep{Zolotov12,SGK18}.  Field haloes do not experience enhanced mass loss of this kind and are thus more robust to the inclusion of baryonic effects in simulations. 

Figure \ref{fig:tbtf} explores the extent to which SFDM may alleviate the TBTF problem in the field.   We show the circular velocities of $M_{\rm h} = 10^{9.5} M_\odot$ haloes (right panel) and $10^{10} M_\odot$ haloes (left) in SFDM (magenta for $m_{22}=2.5$ and cyan for $m_{22} = 0.8$) and CDM (at median concentration, dashed).  These halo masses are similar to those required to match the local counts of $M_\star \simeq 10^6 M_\odot$ dwarf galaxies and are the relevant scale for TBTF comparisons \citep{bullock17}.  The associated scatter in the outer mass distribution due to the variations in $\alpha$ is captured by the shaded regions.  

The star-shaped data points show circular velocities measured at the half-light radii of non-satellite Local Group galaxies as compiled by \citet{garrison-kimmel2014a} \footnote{The data are from \citet{kirby14,hoffman96,simon07,epinat08,frat09,collins13}.}.   Circular velocities of isolated dwarf galaxies measured out to larger radii from \citet{papa16} are shown as pentagons.  The points are color-coded by galaxy stellar mass using the color bar to the right. The color bar reflects galaxy stellar mass.  Note that haloes of galaxies smaller than $M_\star \simeq 5 \times 10^6 M_\odot$ are expected to be only marginally affected by feedback \citep[e.g.,][]{fitts17}.  

The SFDM circular velocity profiles are always below those of CDM in Figure \ref{fig:tbtf} and are generally in better agreement with the data, alleviating the TBTF problem in the field.   In particular, the SFDM models are even able to accommodate the \citet{papa16} data,  which probe the outer radii of galaxies ($\gtrsim 1$ kpc) beyond the stellar half-light radius.   These are regions where hydrodynamic simulations have difficulty lowering the densities of CDM haloes, even with strong feedback \citep{tollet16,fitts17}.  We conclude from this comparison that SFDM provides a possible solution to the TBTF problem with some potential advantages to CDM + feedback models. 

\vspace{1cm}
\subsection{The central density problem and SFDM haloes}

The cusp/core problem in CDM refers to the tendency for measured rotation curves of dark-matter-dominated galaxies to favor fits that imply centrally-cored ($\rho \sim {\rm constant}$) dark matter density profiles as opposed to NFW-like cusps with $\rho \propto r^{-1}$ \citep{McGaugh2001,simon2005,deBlok2008,kuzio-de-naray2008}. A
related issue is that rotation curves indicate lower central
densities than predicted in CDM \citep{alam2002,oman2015}.  A circular velocity scale that is of particular interest for comparison to SFDM models is $\sim 100$  $\mathrm{km\, s^{-1}}$, as these haloes ($M_{\rm h} \mathrm{ \simeq 10^{11}} M_\odot$) are massive enough to have dense soliton cores but small enough that we observe low-density cores in their associated galaxies. 
  
\begin{figure}
\includegraphics[scale=.48,keepaspectratio=true]{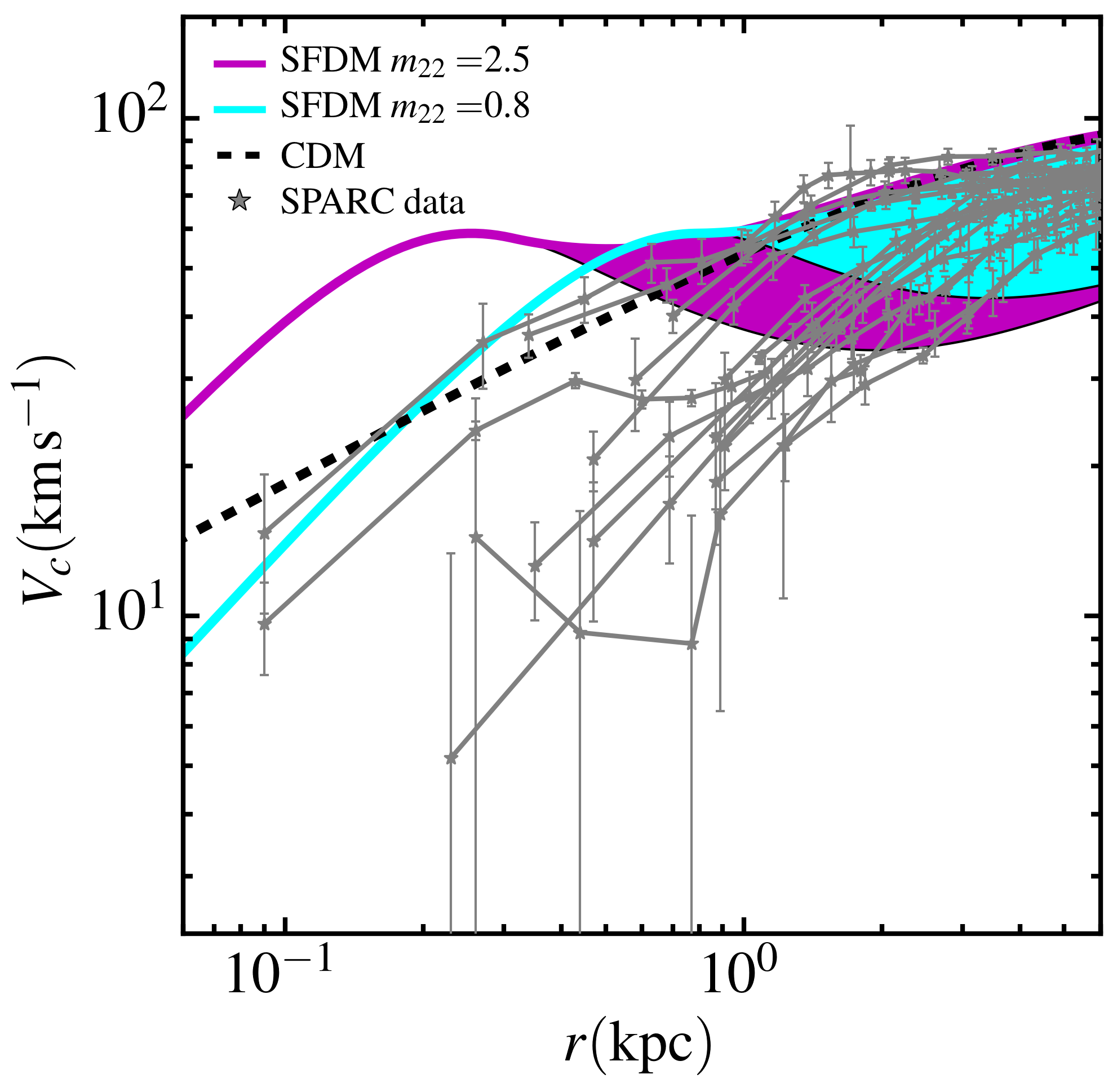} 
\caption{Circular velocities of galaxies from the SPARC database chosen to have asymptotic velocities in the range $\mathrm{80 - 100\,km \, s^{-1}}$ (gray stars connected by thin lines). We include only the dark matter component assuming a stellar mass-to-light ratio of 0.2. These are  compared to predicted rotation curves for $M_{\rm h} \mathrm{ = 10^{11}}M_\odot$ haloes in CDM (dashed) and SFDM (solid). We show SFDM predictions for two scalar field masses, $m_{22}=2.5$ (magenta) and $m_{22}=0.8$ (cyan).  While most of the data at $r>1\,\kpc \sim r_{\rm gal}$ are within the expected scatter for this halo mass in SFDM, the observed rotation velocities at smaller radii are generally lower than those predicted in CDM and SFDM.  The SFDM models are \textit{more} discrepant with the data than CDM at small radii, exacerbating the central-density and cusp/core problem. }
\label{fig:sparc}
\end{figure}

In Figure \ref{fig:sparc}, the gray stars connected by thin gray solid lines show galaxy rotation curves from the SPARC database of galaxy rotation curves \citep{fed16},   specifically the 17 with asymptotic velocities in the range $\mathrm{80 - 100 ~km \,s^{-1}}$.
We have subtracted the gas and disk components from observed rotation curves by assuming a single stellar mass-to-light ratio of 0.2 for all the galaxies at $3.6\,\mu {\rm m}$ (see the discussion in section~5.1 of \citealt{fed16}); we find that changing this value does not modify our conclusions.
For comparison we show the predicted rotation curves for $M_{\rm h} \mathrm{=10^{11}}M_\odot$ haloes in CDM (black dash) and SFDM (solid) with $m_{22}=2.5$ (magenta) and $m_{22}=0.8$ (cyan).  Fourteen of the seventeen galaxies plotted show a decrement in central rotational velocity compared to the CDM expectation, which is characteristic of the cusp/core problem.  Importantly, however, the mismatch at small radii is even worse in the SFDM models owing to the high-density of solitons expected at this circular velocity scale.  
Conversely, the diversity \citep{oman2015} in the data at larger radii ($\sim 2$ kpc) is potentially more easily accommodated in SFDM models.  
We note that significantly smaller scalar field masses ($m_{22}< 0.8$) would better fit the inner regions of the SPARC data at this velocity scale, consistent with the results  of \citet{bernal18}; however, masses this small are ruled out by cosmological constraints \citep{bozek15,sarkar16}.

One important take away from Figure \ref{fig:sparc} is that viable SFDM models do \textit{not} solve the central density problem or cusp/core problem prevalent in $\mathrm{\sim100 ~km \, s^{-1}}$ galaxies.  If anything, SFDM exacerbates the issue (provided Equation \ref{rcmh} is valid).  This suggests that both SFDM and CDM  require some baryonic feedback mechanism to reduce the central densities of haloes at this mass scale.  

\section{Conclusions}

We have presented an analytical formalism for predicting dark matter halo profile structure in Scalar Field Dark Matter (SFDM) models in which the dark matter particle mass is so small ($m \sim 10^{-22}$ eV) that quantum phenomena are important on astrophysically relevant scales ($\sim 1$ kpc). This model, known alternatively as  
Bose-Einstein Condensate Dark Matter,  Ultra-light Dark Matter, Wave Dark Matter, $\psi$DM, and Fuzzy Dark Matter, has gained relevance given its potential for solving some of the small-scale problems in cosmology \citep{hu00,matos00,rob12,hui17,suarez17}.
In view of the demanding spatial resolution required to follow the quantum interference at different scales in SFDM cosmological simulations, our approach offers a useful alternative to describe haloes at all masses and at all radii at $z=0$. Additionally, it can be applied to any scalar field mass $m$. 

As summarized in Section \ref{sec:solitons}, our approach builds upon the  results of \citet{sch14}, who used cosmological simulations to parameterize soliton core sizes $r_c$ and densities as a function of halo virial mass M$_h$.  Importantly, the soliton cores are largest (and lowest density) in the smallest haloes, as might be expected from the uncertainty principle: 
$r_c \sim \lambda_{\rm dB} \propto  m^{-1} v^{-1} \propto m^{-1} \,M_{\rm h}^{-1/3}$.  
The soliton regions transition to an outer profile that mirrors CDM expectations in the regime where quantum pressure is negligible.  We parameterize the transition radius $r_\alpha = \alpha \, r_c$ with a range of $\alpha$ values that are plausible and that characterize the expected range where the soliton transitions from a turbulent region to the dust-like CDM halo (see Section \ref{sec:envelopes}). 
Assuming the validity of Equation \ref{rcmh},  the central regions of SFDM haloes will be dominated by solitons  that are denser than the same halo in CDM at masses above M$_h \sim 10^{10.5}$ M$_\odot$ (see, e.g., Figure \ref{rho_vc}). 

Using our formalism, we consider the smallest SFDM masses ($m = 0.8-2.5 \times 10^{-22}$ eV) that are still consistent with large-scale structure constraints, as such models will result in manifestations of quantum pressure (solitons) on the largest astrophysical scales. In such models, dwarf-size dark matter halos (M$_h \sim 10^{10}$ M$_\odot$) have lower central densities than their CDM counterparts, alleviating the Too Big to Fail problem in a regime where feedback is less effective (see Figure \ref{fig:tbtf}).  However, the cusp/core problem seen in more massive (M$_h \sim 10^{11} $M$_\odot$) dwarf galaxies is exacerbated.  We use rotation curves of $\sim 100$ km s$^{-1}$ galaxies from the SPARC database to show that SFDM halos are denser at the radii where rotation curves rise than even cuspy NFW haloes. This makes the low central densities observed in dark-matter dominated galaxies even harder to understand in SFDM models than in CDM (Figure \ref{fig:sparc}).   For Milky Way mass haloes, SFDM has a soliton density well above the CDM expectation at $\sim 100$ pc, some $\sim 10$ times denser than the equivalent NFW profile at those radii (see also \citealt{bar18}).  

Our results motivate future large-scale SFDM simulations in order to confirm the predicted soliton scalings with halo mass and to explore the expected scatter in transition radii from soliton to NFW-like envelopes.  Updated scalings (e.g., an updated form of Equation \ref{rcmh}) can be easily accommodated in our approach.  They also motivate the need to incorporate hydrodynamics and star-formation feedback into SFDM simulations in order to determine whether or not baryonic processes can alleviate the core-density tension for SFDM highlighted in Figure \ref{fig:sparc}.  If not, then one of the classic small-scale problems that originally motivated SFDM may be its undoing.

\section*{Acknowledgments}
V.H.R. acknowledges support from UC-MEXUS and CONACyT through the postdoctoral fellowship.  This work was also supported by NSF grants AST-1518291 and PHY-1620638 and by NASA through HST theory grants (programs AR-13921, AR-13888, and AR-14282) awarded by the Space Telescope Science Institute (STScI), which is operated by AURA, Inc., under NASA contract NAS5-26555. MBK also acknowledges support from NSF grant AST-1517226 and CAREER grant AST-1752913 and from NASA grants NNX17AG29G and HST-AR-13896, HST-AR-14554, HST-AR-15006, HST-GO-12914, and HST-GO-14191 from STScI.

\section*{Appendix: SFDM halo profile with oscillations}

Beyond the inner soliton radius, the density field exhibits a homogeneous and isotropic turbulence that is caused by quantum interference. \cite{mocz17} found the presence of a characteristic scale ($\sim$ twice the soliton radius): at the scale of the soliton, the de Broglie wavelengths of the dark matter particles overlap with the same phase and interfere constructively.

At the halo center, a large number of particles remain bounded and possess comparable velocities (having a single phase leading to coherent interference), implying a large increase in the local density within the soliton. Beyond the central soliton, the phase of the field gradually loses coherence due to (i) the flow of incoming and outgoing particles with different velocities (phases) and (ii) the decreasing mass density. These two effects imply that for larger radii, the average interference-induced increase of local density will be much less pronounced and the average effect of the turbulent medium might be approximated as a small overdensity of the time-averaged local density. 

It is interesting to explore the potentially observable consequences of this interference.  We explore this possibility by considering a perturbation to the outer NFW profile using $\epsilon$, a constant parameter that regulates the contribution of the perturbation (0$\leq \lvert \epsilon \rvert \leq$1):
\begin{equation}
\rho^{\rm osc}_{\rm NFW}(r )=\rho_{\rm NFW}(r)[1+\epsilon \,h(r)]\, \hspace{0.25cm}  {\rm for }\; r_{\alpha} \leq r\,.
\label{nfwosc}
\end{equation}
Here $h(r)$ is a normalized function ($0 \leq \lvert h (r) \rvert \leq 1$) that defines the fluctuations and is derived below. 

We will assume that, on average, there is a single dominant wavelength that modulates the oscillations. A natural choice is the characteristic scale associated with each individual halo, the soliton size (=$2\,r_\alpha$); the modulating function describing the perturbations is taken as a periodic normalized soliton profile, such that in one period is given by $g(r)$, defined by:
\begin{equation}
g(r )=\frac{\rho_{\rm sol}(r-r_{\alpha}+ n(2r_\alpha))-\rho_{\rm sol}( (n+1)r_\alpha)}{\rho_c-\rho_{\rm sol}(r_\alpha)}, 0\leq r \leq 2r_\alpha
\end{equation}
with $n=r\,{\rm mod}\,(4\,r_\alpha)$. With this definition, we now construct the periodic function $w(r)$ for all radii as:
\begin{equation}
        w(r )=
        \begin{cases}
            g(r-4nr_\alpha)&n(4\,r_\alpha)\leq r \leq (2n+1)2\,r_\alpha \\
            g(r-2r_\alpha-4\,n\,r_\alpha) &(2n+1)2\,r_\alpha \leq r \leq (n+1)4\,r_\alpha.
        \end{cases}
    \end{equation}  
As the correction is valid for $r\geq r_\alpha$ (outside of the inner soliton), our perturbation function is $h( r):= w(r-r_\alpha)$. 

The SFDM profile with oscillations is given by:
\begin{equation}
        \rho^{\rm osc}(r )=
        \begin{cases}
        \displaystyle
            \rho_{\rm sol}( r)&0\leq r \leq r_\alpha \\
            \rho^{\rm osc}_{\rm NFW}(r )= \dfrac{\rho^{\rm osc}_{\rm s} (1+\epsilon \,h(r))}{(r/r^{\rm osc}_{\rm s})(1+r/r^{\rm osc}_{\rm s})^2} &r_\alpha \leq r_{\rm vir}
        \end{cases}
    \label{becosc}
    \end{equation}
We apply the eqs.~(\ref{xxx}--\ref{yyy}), now using eq. (\ref{becosc}) for the density, to obtain the corresponding NFW profile for the outer tail (numerically solving the system for the new scale $r^{\rm osc}_s$).

\begin{figure*}
\centering 
\hspace{-.2cm}\includegraphics[scale=.65,keepaspectratio=true]{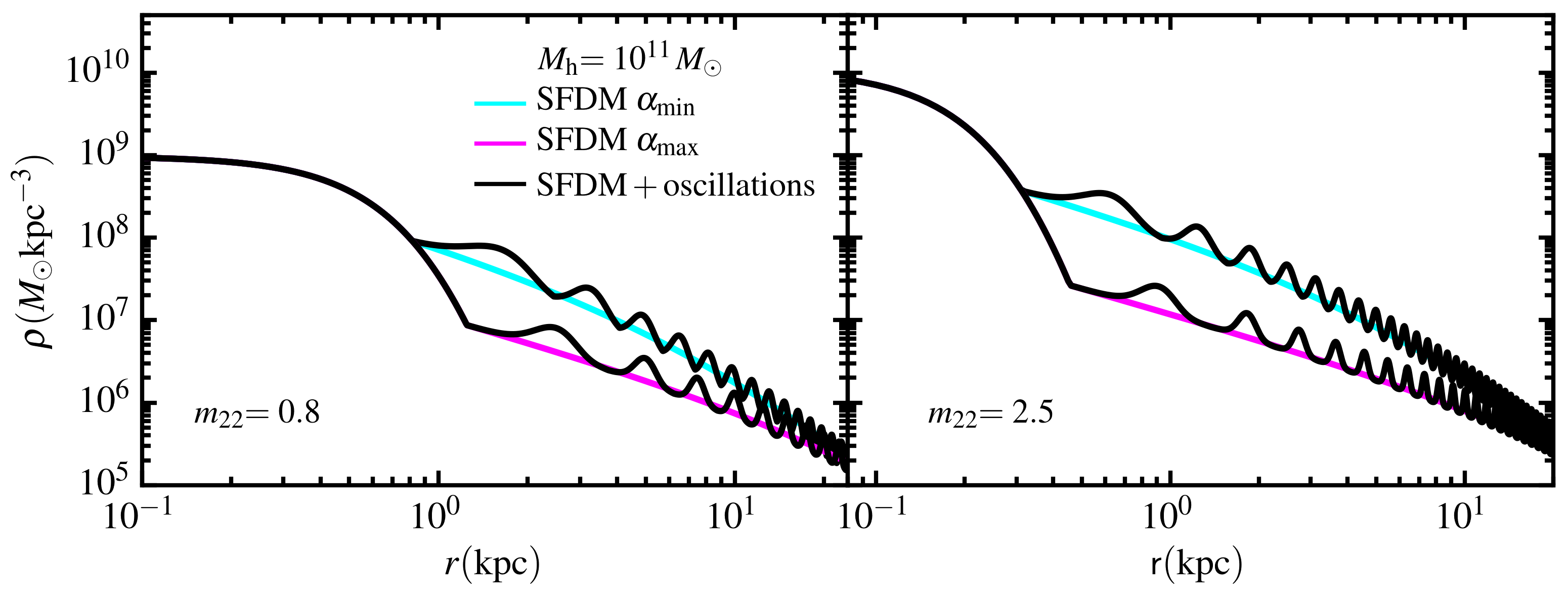} \\
\includegraphics[scale=.65,keepaspectratio=true]{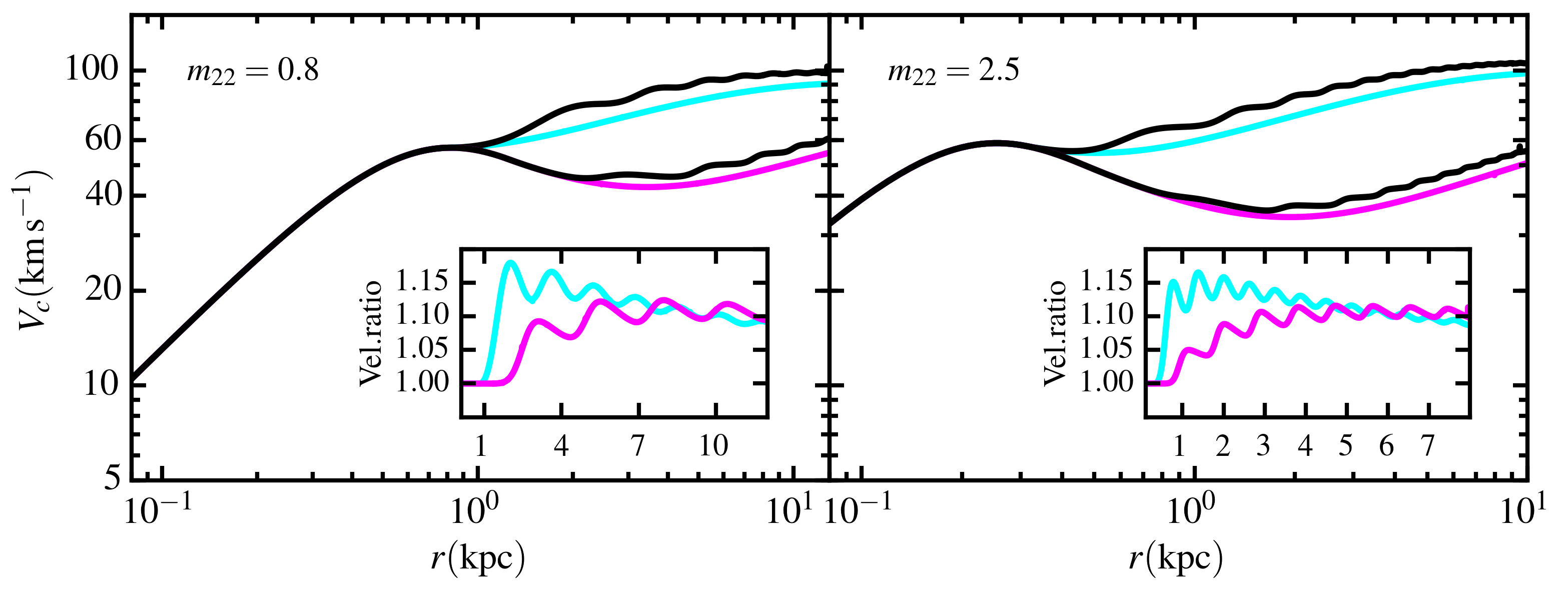} 
 \caption{SFDM density profile (top panel) and circular velocity (bottom panel) for a halo mass of $M_{\rm h}=10^{11}M_{\odot}$, similar to haloes hosting low-surface brightness galaxies. We show two particle masses as labeled. We compare the smooth profiles for the maximum soliton (magenta), the minimum soliton (cyan), and their respective corrections (black). We observe the presence of wiggles in the density profile, modeling the net effect of the fluctuating medium according to eq. \ref{becosc}. In our derivation for the SFDM with oscillations, we are considering the oscillating field is mostly described by the dominant mode with characteristic scale equal to the soliton size ($=2 r_\alpha$). These over densities could trap some gas or stars and might enhance star formation in these regions, in the rotation curve, these soliton-size oscillations would be seen as wiggles in the profile, smaller solitons imply more compact ripples and would require high spatial resolution to observe the oscillation. The inset in the bottom figures compares the ratio between the SFDM profiles with and without oscillations, we observe that there is a $10\%-15\%$ difference in the amplitude, the predicted ripples might be large enough to be observable, perhaps in the form of diffuse gas shells or rings.}
 \label{fig8}
\end{figure*}

In Fig. \ref{fig8}, we compare the SFDM profile with and without fluctuations for a halo with mass $M_{\rm h}=10^{11}\rm M_{\odot}$, typical for low-surface brightness galaxies,  and with $\epsilon=1$ to illustrate the effect of the correction to the smooth NFW profile. We notice that even for this maximal value of $\epsilon$, the density profile displays only small density oscillations that become less evident for larger radii. The largest difference is in the first few soliton-size overdensities, where densities reach $\sim10\%-15\%$ larger than the smooth NFW-tail fit with no oscillations. These overdensites imply the existence of oscillations (or "wiggles") in the circular velocity that may be probed observationally, as seen in the bottom panels of Fig. \ref{fig8}. Smaller solitons result in oscillations with a smaller period (length), as can be seen in the figure; these may require observations with very high spatial resolution to be detectable.

While oscillations in the circular velocity profile may be an observable signature of a quantum nature of dark matter, we note there are at least two factors that should be taken into account when searching for oscillations in rotation curves. First, baryonic matter may affect the distribution of the inner halo via, e.g., baryonic contraction; this could change the soliton size and consequently the characteristic length and amplitude of the oscillations. Second, we are only assuming a dominant wavelength to model the interference; in a more realistic scenario, we should include the contribution of the different phases giving rise to the turbulent medium. Such a study would require a mode decomposition, and $h(r)$ would become a more complicated function that would include the sum of all the different modes (a similar approach following this idea was given in \citealt{rob13,martinez15,bernal17,bernal18}). Depending on the individual coefficients for the new modulating function, the oscillations may not appear at exact periodic intervals, as is assumed by our choice of $h(r)$. In fact, if modes add coherently, they could increase the fluctuation amplitude in some regions to an observable level, possibly resulting in ripples in the circular velocities of low-density gas or the kinematics of halo stars. Current and future {\em Gaia} data releases may be able to explore this scenario.  

We find that individual \textit{single-frequency} fluctuations become small in massive haloes (with small soliton sizes), making them indistinguishable from a smooth distribution at very large radii. Observationally, the first few overdensities closest to the soliton are the best place to search for this unique feature of the SFDM model.  Though limited, our phenomenological approach could be extended to more complicated models of the fluctuating field and may have some implications for the diversity of rotation curves in spiral galaxies \citep{oman15}. Once large-scale SFDM simulations are available, we may be able to obtain statistical halo properties and calibrate the modulating function presented in this appendix.

\bibliography{mybib}

\bsp
\label{lastpage}
\end{document}